\documentclass[a4paper,11pt]{article}
\usepackage{graphicx}
\usepackage{makeidx}
\usepackage{natbib}
\newcommand\eq[1] {(\ref{#1})}

\newcommand{\bfm}[1]{\mbox{\boldmath ${#1}$}}

\newcommand{\beqa}{\begin{eqnarray}}
\newcommand{\eeqa}[1]{\label{#1}\end{eqnarray}}
\newcommand{\bequ}{\begin{equation}}
\newcommand{\eequ}[1]{\label{#1}\end{equation}}

\newcommand{\ov}[1]{\overline{#1}}

\newcommand{\Gve}{\varepsilon}

\newcommand{\Gl}{\lambda}

\newcommand{\Gs}{\sigma}

\newcommand{\Go}{\omega}

\newcommand{\GF}{\Phi}

\newcommand{\GT}{\Theta}
\newcommand{\GS}{\Sigma}

\newcommand{\BGs}{\bfm\sigma}

\newcommand{\BGj}{\bfm\tau}

\newcommand{\BGS}{\bfm\Sigma}

\newcommand{\CO}{{\cal O}}

\newcommand{\BCA}{{\bfm{\cal A}}}
\newcommand{\BCB}{{\bfm{\cal B}}}

\newcommand{\BCM}{{\bfm{\cal M}}}

\newcommand{\BCR}{{\bfm{\cal R}}}

\newcommand{\BCT}{{\bfm{\cal T}}}
\newcommand{\BCU}{{\bfm{\cal U}}}

\def\Bg{{\bf g}}
\def\Bh{{\bf h}}

\def\Bp{{\bf p}}

\def\Bt{{\bf t}}
\def\Bu{{\bf u}}
\def\Bv{{\bf v}}
\def\Bw{{\bf w}}

\def\Bz{{\bf z}}
\def\BA{{\bf A}}
\def\BB{{\bf B}}

\def\BH{{\bf H}}

\def\BJ{{\bf J}}
\def\BK{{\bf K}}
\def\BL{{\bf L}}

\def\BQ{{\bf Q}}
\def\BR{{\bf R}}

\def\BT{{\bf T}}
\def\BU{{\bf U}}

\def\BY{{\bf Y}}

\newcommand{\beq}{\begin{equation}}
\newcommand{\eeq}{\end{equation}}
\newcommand{\overliner}{\begin{eqnarray}}
\newcommand{\earr}{\end{eqnarray}}
\newcommand{\beqn}{\begin{equation*}}
\newcommand{\eeqn}{\end{equation*}}
\newcommand{\overlinern}{\begin{eqnarray*}}
\newcommand{\earrn}{\end{eqnarray*}}

\newcommand{\fr}{\frac}

\begin{document}

\centerline{\Large \textbf{Stroh formalism in analysis of  skew-symmetric }}

\centerline{\Large \textbf{ and symmetric  weight functions for interfacial cracks}  }

\begin{center}
L. Morini$^{(1)}$, E. Radi$^{(1)}$, A.B. Movchan$^{(2)}$, and N.V. Movchan$^{(2)}$\\
\end{center}

\centerline{$^{(1)}$\emph{Dipartimento di Scienze e Metodi dell'Ingegneria, Universit\'a di Modena e Reggio Emilia,}}

\centerline{\emph{Via Amendola 2, 42100, Reggio Emilia, Italy}}

\centerline{$^{(2)}$\emph{Department of Mathematical Sciences, University of Liverpool,}}

\centerline{\emph{Liverpool L69 3BX, U.K.}}

\begin{abstract}
The focus of the article is on analysis of skew-symmetric weight matrix  functions for interfacial cracks in two dimensional anisotropic solids. It is shown that the Stroh formalism proves to be an efficient approach to this challenging task.
Conventionally, the weight functions, both symmetric and skew-symmetric, can be identified as a non-trivial singular solutions of the homogeneous boundary value problem for a solid with a crack. For a semi-infinite crack, the problem can be reduced to  solving a matrix Wiener-Hopf functional equation. Instead, the Stroh matrix representation of displacements and tractions, combined with a Riemann-Hilbert formulation, is used  to obtain  an algebraic eigenvalue problem, that is solved in a closed form. The proposed general method is applied to the case of a quasi-static semi-infinite crack propagation between two dissimilar orthotropic media: explicit expressions for the weight matrix functions are evaluated and then used in the computation of complex stress intensity factor corresponding to an asymmetric load  
 acting on the crack faces.\\

%In this article we study the vector problem of a 2D semi-infinite static crack propagation at the interface between two dissimilar anisotropic elastic materials, loaded by a general asymmetric system of forces acting on the crack faces. Symmetric and skew-symmetric weight functions matrices, required for the evaluation of stress intensity factors, are derived via the direct solution of the homogeneous Riemann-Hilbert problem associated to a semi-infinite interfacial crack, obtained by means of Stroh matricial representation of displacement fields and tractions at the interface. The proposed general method is applied to the case of a static semi-infinite crack propagation between two different orthotropic media: explicit expressions for the weight functions matrices are evaluated and then used in the computation of complex stress intensity factor corresponding to an illustrative example of asymmetric system of forces acting on the crack faces.\\

\end{abstract}
%\tableofcontents
\section{Introduction}
Evaluation of coefficients in asymptotic representations of displacements and stress fields represents an important issue for addressing vector problems of crack propagation in elastic materials \citep{BercVel1,MishKuhn1}. The explicit derivation of weight functions is fundamental for the evaluation of stress intensity factors corresponding to a general distribution of forces acting on the crack faces, as well as for the calculation of higher order coefficients in the asymptotic expressions of the fields. The latter are used in asymptotic models of incremental crack growth, and hence are essential for evaluation of the crack path and analysis of fracture stability. In the work by Bueckner \citep{Bueck1,Bueck2}, weight functions for several types of cracks in homogeneous elastic media, both in two and three dimensions, have been defined.
%, and stress intensity factors corresponding to several loading configurations applied on the crack boundaries are estimated. 

 In this paper, the term "symmetric" load is associated to forces of the same magnitude applied on both crack faces in opposite directions, while the load generated by forces acting on both crack faces in the same direction is called "skew-symmetric" or "anti-symmetric". For cracks 
 %placed 
 in homogeneous elastic materials, in the two-dimensional setting the skew-symmetric loading does not contribute to stress intensity factors, whereas it becomes relevant and it must to be accounted  for in three-dimensional solids \citep{Bueck1,MeadKeer1}.
 %, as well as for interfacial cracks  between dissimilar media in two dimensions. 
 The situation is different when the crack is placed at the interface between two dissimilar elastic materials: 
 %in this case, 
 even 
 %considering plane 
 for two-dimensional problems %the stress fields oscillate and also 
 the skew-symmetric loads generate a non-zero contribution to stress intensity factors \citep{BercVel1,LazLebl1, PiccMish1}. In particular, for Mode III interfacial cracks % where 
 the stress components do not oscillate, but  a non-vanishing skew-symmetric component of the weight function still has to be accounted  \citep{PiccMish1,PiccMish4}. In the case of isotropic media, weight functions for semi-infinite % interfacial 
 cracks can be defined as singular non-trivial solutions of the homogeneous boundary problem with zero tractions on the crack faces but unbounded elastic energy \citep{Wilmov1}. For interfacial cracks between dissimilar isotropic elastic media, the weight functions  are well discussed in literature \citep{LazLebl1,PiccMish1,PiccMish4}. The problem is generally reduced to a functional equation of Wiener-Hopf type, and its solution gives the symmetric weight function matrix \citep{Ant1}, while the skew-symmetric component is obtained by the construction of the corresponding full-field singular solution of the elasticity boundary value problem discussed in \citet{PiccMish1}. For interfacial cracks between anisotropic dissimilar elastic media, although weight functions have been derived by \citet{Gao2,Gao1} and \citet{MaChen1}, the results on skew-symmetric weight functions are not readily available.
 
 In this article we illustrate a general procedure for the calculation of symmetric and skew-symmetric weight functions matrices for semi-infinite two-dimensional interfacial crack problems in anisotropic elastic bi-materials. It is shown that the challenging analysis of the matrix functional Wiener-Hopf equation can be replaced by solving a matrix eigenvalue problem deduced via an equivalent formulation, based on Stroh representation of displacement and stress fields \citep{Stroh1}. By means of this new approach, general expressions for weight functions matrices, valid for plane interfacial cracks problems between any anisotropic media, are obtained. This general result is reported and discussed in details in Section (\ref{weight_func}). 
 
 Section (\ref{ortho}) illustrates the proposed method for the case of a semi-infinite two-dimensional stationary crack between two dissimilar orthotropic materials under plane stress deformation \citep{Suo1}, corresponding to a general non-symmetric load. Stroh representations for displacements and stress corresponding to this problem are explicitly calculated and used for deriving symmetric and skew-symmetric weight functions matrices. In Section (\ref{example}) both symmetric and skew-symmetric weight functions are utilized together with Betti formula in order to evaluate complex stress intensity factor for an interfacial crack in orthotropic bi-material subject to an asymmetric loading configuration. In the particular case of isotropic media, the obtained result is consistent with the stress intensity factor derived for the non-symmetric  distribution of forces obtained by 
 % applied on the faces of a crack placed at the interface between two isotropic media (see 
  \citet{PiccMish1}.
 
 Finally, in Appendix A, the evaluated skew-symmetric weight function is compared to those calculated by the full field singular solution of the plane elasticity problem, following the approach illustrated in \citet{PiccMish1}. The perfect agreement detected between the expressions derived by means of two alternative formulations represents an important benchmarking for the correctness of the obtained results.
\section{Interfacial cracks: preliminary results}
\label{interface}
\begin{figure}[htbp]
\centering
\includegraphics[width=9cm,height=5.6cm]{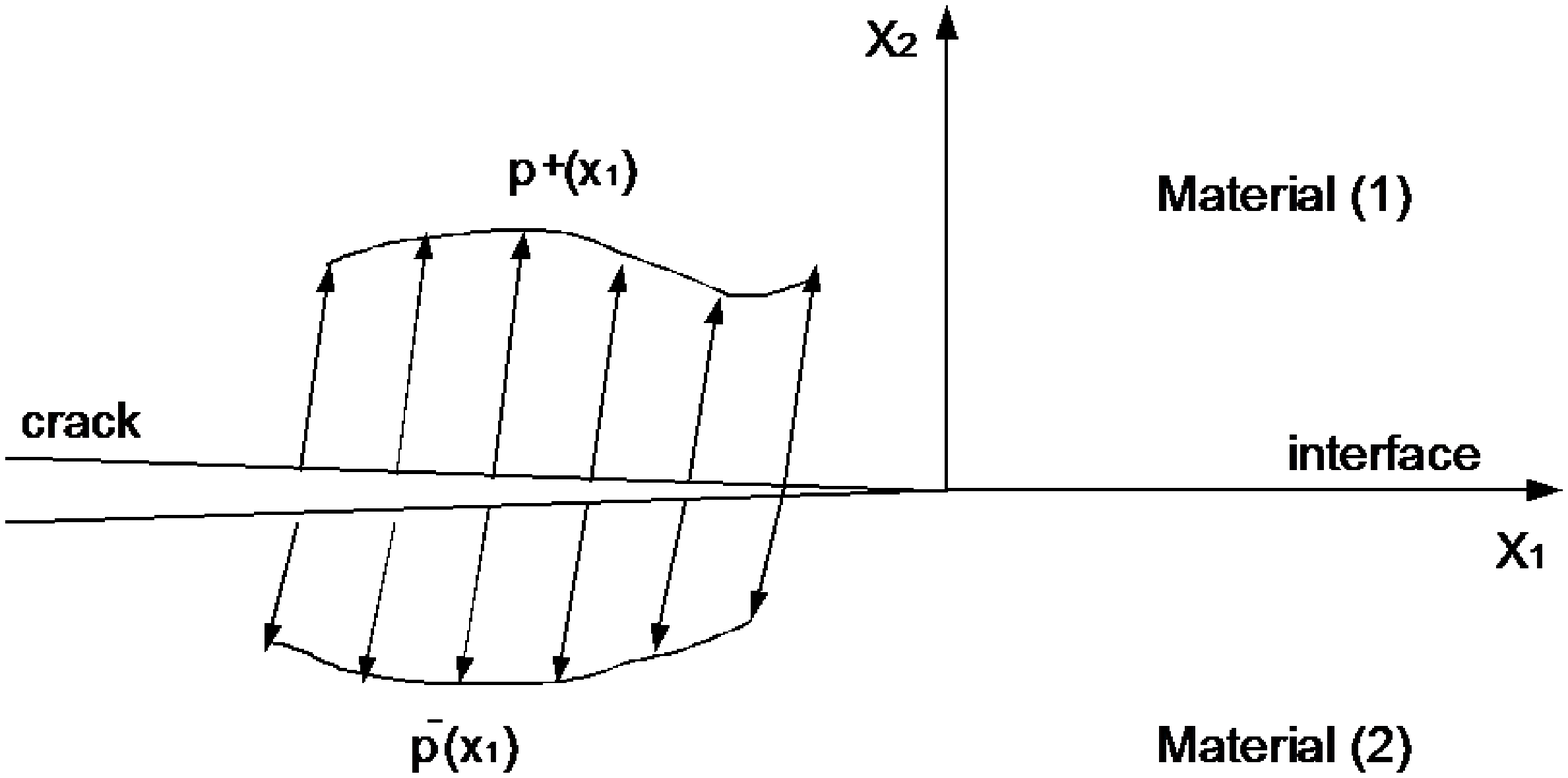}
\caption[geom_frac]{\footnotesize {Geometry of the model.}}
\label{geom_frac}
\end{figure}

%We start introducing 
Here we introduce the main notations and the mathematical framework of the model. 
%The problem investigated is the propagation 
Consider a quasi-static advance of a semi-infinite plane crack between two dissimilar anisotropic elastic materials with asymmetric loading applied to the crack faces, the geometry of the system is shown in Fig.(\ref{geom_frac}).

The loading is defined via  tractions acting upon the crack faces. Considering a Cartesian coordinate system with the origin at the crack tip, (see Fig.(\ref{geom_frac})), traction components behind the crack front are then defined as follows:
\beq
\Gs_{2j}^{\pm}(x_{1},0^{\pm})=p_{j}^{\pm}(x_{1})\quad \textrm{for} \quad x_{1}<0 \quad j=1,2,
\eequ{p}
where $p_{j}^{\pm}(x_{1})$ are given functions. 

Since the load is assumed to be self-balanced, its resultant force and moment vectors are equal to zero. Moreover, we assume that forces are applied outside a neighborhood of  the crack tip and vanish at the infinity. The body forces are assumed to be zero. Symmetric and skew-symmetric parts of the loading are given by the following expressions: 
\beq
\left\langle p_{j}\right\rangle(x_{1})=\fr{1}{2}\left(p_{j}^{+}(x_{1})+p_{j}^{-}(x_{1})\right),\quad [p_{j}](x_{1})=p_{j}^{+}(x_{1})-p_{j}^{-}(x_{1}),\quad j=1,2,
\eequ{p_symskew}

 The solutions in form of functions which vanish at the infinity and possess finite elastic energy are sought. Expressions for the stress field and displacements for a semi-infinite interfacial crack in anisotropic bi-materials have been obtained by means of Stroh formalism \citep{Stroh1} by \citet{Suo1}. This approach to the physical problem of the crack, that is used for the derivation of weight function matrix in the paper, is reported in Section (\ref{Stroh_int}). 

In Section (\ref{weight_betti}) the weight functions are defined as non-trivial singular solutions of the homogeneous elasticity problem for interfacial cracks with zero tractions on the crack faces but unbounded elastic energy, following the approach of \citet{PiccMish1}. 

Section (\ref{Betti_id}) reports the application of the Betti formula to physical fields and weight functions and the derivation of the fundamental identity in the Fourier space, discussed in details in \citet{Wilmov1,PiccMish3} and \citet{PiccMish2}. This integral identity will be used in the paper for the evaluation of coefficients in the asymptotics of the stress fields near the crack tip by means of a procedure based on weight functions theory \citep{PiccMish1}. 

The asymptotic representations of physical displacements and stress fields near the crack tip, including the stress intensity factor as the coefficient of the first term \citep{PiccMish3,PiccMish1}, are introduced in Section (\ref{asym_k}).

\subsection{Stroh formalism in analysis of interfacial cracks}
\label{Stroh_int}
 Physical displacements and stress fields for an intefacial crack between two different anisotropic materials can be derived by means of 
 %alternative 
 the Lekhnitskii or Stroh approaches \citep{Lekh1,Stroh1,Suo1}. Following the article by Suo \citep{Suo1}, we introduce the stress vectors ${\bf t}_j = (\Gs_{1j}, \Gs_{2j})^T$ together with the displacement $\Bu = (u_1, u_2)^T$. The constitutive relations for both the elastic media occupying respectively the upper and lower half-planes can be written using the Stroh formulation \citep{Stroh1}:
\begin{eqnarray}
\Bt_1 & = & \BQ \Bu_{,1} + \BR \Bu_{,2}   \\
\Bt_2 & = & \BR^T {\Bu}_{,1} + \BT \Bu_{,2},
\end{eqnarray}
where the matrices $\BQ, \BR$ and $\BT$ depend on the material constants. A semi-infinite static crack placed at the interface between the two materials is considered, as it is illustrated in Fig.(\ref{geom_frac}). Accordingly, the derivative of the displacement $\Bu_{,1}(x_{1},x_{2})$ and the traction $\BGj(x_{1},x_{2}) = \Bt_2(x_{1},x_{2})$ can be written as
 \beq
 \BGj(x_{1},x_{2}) = \BB \Bg(\Bz) + \ov{\BB \Bg(\Bz)}
 \eequ{eq1}
and
 \beq
 \Bu_{,1} (x_{1},x_{2}) = \BA \Bg(\Bz) + \ov{\BA \Bg(\Bz)},
 \eequ{eq2}
 where $\BA$ and $\BB$ are constant matrices, $\Bg(\Bz)$ is an analytic vector function with components $ g_j(x_{1}+ \mu_j x_{2})$, and $\mu_j$ are complex numbers with positive imaginary parts.
 According to 
 %the property enunciated by 
 \citet{Suo1}, if $g_{j}(z_{j})$ is an analytic function of $z_{j}=x_{1}+ \mu_j x_{2}$ in the upper half-plane (or in the lower half-plane) for one $\mu_j$, where $\mu_j$ is a complex number with positive imaginary parts, it is analytic for any  $\mu_j$. On the basis of this property, here and in the text that follows, we reduce the analysis to a single complex variable.  The connection between the elements of $\BA$, $\BB$ and $\BQ, \BR, \BT$ is given by the following relations (see \citet{Ting1}, pages 170, 171):
\beq
(Q_{ik}+(R_{ik}+R_{ki})\mu_{j}+T_{ik}\mu_{j}^{2})A_{kj} = 0 
\eequ{eq3}
\beq
B_{ij}=(R_{ki} + \mu_{j} T_{ik})A_{kj}                                                                                                                                                                          \eequ{eq4}  
Thus, each column of $\BA$ is a non-trivial solution of the eigensystem (\ref{eq3}), while the eigenvalues $\mu_j$  are roots of the characteristic equation: 
\beq
|Q_{ik}+(R_{ik}+R_{ki})\mu_{j}+T_{ik}\mu_{j}^{2}| = 0 
\eequ{eq5}                                                                                                                                                                                     
 In turn, for each of the two phases we introduce the Hermitian matrix $\BY = i \BA \BB^{-1}$, which will be used in the text below. Further in the text, we shall use the superscripts $^{(1)}$  and  $^{(2)}$ to denote the quantities related to the upper and lower half-planes, respectively.
 
 The expression \eq{eq1}, in the limit $x_{2}\rightarrow 0^{\pm}$, leads to a non-homogeneous Riemann-Hilbert problem:
\beq 
 \Bh^+(x_1) - \Bh^-(x_1) = \BGj(x_{1}), \quad x_1\in\mbox{R},
\eequ{eq5} 
 
 A crack advancing  along the negative semi-axis $x_{1} < 0$ is considered, the traction-free condition is imposed for $x_{1}<0$, while the continuity of the tractions at the interface is assumed ahead the crack front. The following equations are satisfied on the real axis \citep{Suo1}:
 \begin{eqnarray}
 \Bh^+(x_{1}) + \overline{\BH}^{-1}\BH\Bh^-(x_{1}) & = & \BGj(x_{1})\quad \textrm{for}\quad x_{1}>0\\
 \Bh^+(x_{1}) + \overline{\BH}^{-1}\BH\Bh^-(x_{1}) & = & 0 \qquad \textrm{for}\qquad x_{1}<0 \label{RH_hom1}
 \end{eqnarray}
 
 The detailed analysis of this problem is included in \citet{Suo1,SuoKuo1}, and it shows that the stress and displacement fields do not have oscillation near the crack tip for the case when the matrix $\BH = \BY^{(1)} + \ov{\BY}^{(2)}$ is real, otherwise the stress and displacement are characterized by the oscillatory behavior near the crack tip. The branch cut for the function $\Bh(z)$ is assumed to be along the crack line $x_{1}<0$. Assuming that the stress vanish at the infinity, the solution of the homogeneous Riemann-Hilbert problem (\ref{RH_hom1}) is $\Bh(z)=\Bw z^{-\fr{1}{2}+i\varepsilon}$, where $\Bw$ is solution of the following eigenvalues problem \citep{GaoAbbu1,Suo1,SuoKuo1}:
\beq
\ov{\BH} \Bw = e^{2 \pi \Gve} \BH \Bw.
\eequ{eigens} 

The traction  $\BGj$ ahead of the crack has the form:
\beq
\BGj(x_{1}) = \fr{1}{\sqrt{2 \pi x_{1}}} \mbox{Re} ~ \Bigg(  K x_{1}^{i \Gve}  \Bw \Bigg).
\eequ{tract}
where $K=K_{I}+iK_{II}$ is the complex stress intensity factor including both mode I and mode II contributions to the traction, $\Gve$ is the bi-material parameter (a real non-dimensional number measuring an aspect of elastic dissimilarity of the two materials), and $\Bw$ is the eigenvector obtained from \eq{eigens}.

 For the case of plane strain load, it has been shown in Suo \citep{Suo1,SuoKuo1} that the displacement jump $[\Bu]$ across a semi-infinite crack running along the negative semi-axis $x_{1} < 0$ is given in the form
\beq
 [\Bu](x_{1}) =  \Bigg(  \fr{2(-x_{1})}{ \pi} \Bigg)^{1/2}  \fr{(\BH + \ov{\BH})}{\cosh \pi \Gve} \mbox{Re}~\Bigg(  \fr{K(-x_{1})^{i \Gve} \Bw }{1 + 2 i \Gve}     \Bigg)  
\eequ{jump}
%where $r$ is the distance from the crack tip.
\subsection{Weight functions definition}
\label{weight_betti}
 Following the theory developed by Willis and Movchan \citep{Wilmov1}, we define a 
 %singular displacement 
 vector function $\BU=(U_1,U_2)^{T}$ as the singular solution of the elasticity problem with zero traction on the faces where the crack is placed along the positive semi-axis $x_{1}>0$. The traces of these functions on the plane containing the crack are known as the weight functions \citep{PiccMish1}, and notations $[\BU]$ and $\langle \BU \rangle $ will be used in the paper to denote symmetric and skew-symmetric weight functions respectively:
\begin{eqnarray} 
[\BU](x_{1})                & = & \BU(x_{1},x_{2}=0^{+})-\BU(x_{1},x_{2}=0^{-})\label{Usymd}\\
\langle \BU \rangle (x_{1}) & = & \fr{1}{2}(\BU(x_{1},x_{2}=0^{+})+\BU(x_{1},x_{2}=0^{-}))\label{Uskewd}
\end{eqnarray} 

   The traction vector $\BGS=(\GS_1,\GS_2)^{T}$ associated to the singular solutions is continuous on the plane containing the crack (see \citet{PiccMish2,PiccMish1}) and vanishes for $x_{1}<0$ (homogeneous boundary conditions are imposed). In practice, for singular solutions we impose traction-free condition for $x_{1}>0$ and traction continuity at the interface for $x_{1}<0$:
\begin{eqnarray}
\Bh^+(x_{1}) + \overline{\BH}^{-1}\BH\Bh^-(x_{1}) & = & 0 \qquad \textrm{for}\qquad x_{1}>0\label{RH_nohom2}\\
\Bh^+(x_{1}) + \overline{\BH}^{-1}\BH\Bh^-(x_{1}) & = & \BGS(x_{1})\quad \textrm{for}\quad x_{1}<0 \label{RH_hom2}
\end{eqnarray}

 It is important to note  that the domain of singular solutions and then of weight functions is different from the domain of the physical solution, defined in previous Section, where the crack is placed along the negative semi-axis. For the singular solution, the branch cut for $\Bh(x_{1})$ is chosen to be along the line $x_{1}>0$. %and introducing the ansatz 
 Using the representation $\Bh(z)=\Bv z^{-\fr{3}{2}+i\varepsilon}$ we reduce  (\ref{RH_hom2}) %is reduced 
 to the eigenvalues problem:
\beq
\ov{\BH} \Bv = e^{-2 \pi \Gve} \BH \Bv.
\eequ{eigens_sing}
 The bi-material matrix $\BH$ is %assumed to be definite positive and hermitian, then 
 Hermitian, and the eigenvector corresponding to the eigenvalue $-\Gve$ is $\Bv=\overline{\Bw}$ \citep{Suo1}. The singular traction $\BGS$ 
 %assumes 
 is sought in the form:
\beq
\BGS=\fr{(-x_{1})^{-\fr{3}{2}}}{\sqrt{2\pi}}\textrm{Re}\left(C\ov{\Bw}(-x_{1})^{i\varepsilon}\right),
\eequ{sigma}
where $C=C_{I}+iC_{II}$ is a complex constant representing both mode I and mode II contributions to the traction \eq{sigma}, as it is for the complex stress intensity factor $K$ with regard to the physical traction (see eq.\eq{tract}). The shear and normal opening modes for plane stress and plane strain problems are coupled. Two linearly independent vectors $\BGS$ and associated weight functions can be identified similar to  \citet{PiccMish3,PiccMish1}.
\subsection{Fundamental Betti identity}
\label{Betti_id}
%In Sections (\ref{Stroh_int}) physical diplacements $\Bu = (u_1, u_2)^T$ and traction fields $\BCT = (\CT_1, \CT_2)^T$ ahead of the crack tip are introduced by means of Stroh formalism, while in Section (\ref{weight_betti}) 
%singular displacements $\BU=[U_1,U_2]^{T}$ and the corresponding tractions $\BGS=[\GS_1,\GS_2]^{T}$ are defined. As just
As discussed above, $\BU$ is discontinuous along the positive semi-axis $x_{1}>0$, whereas $\Bu$ is discontinuous along the positive semi-axis $x_{1}<0$. $\BGS$ is zero for $x_{1}>0$, whereas $\BGj$ is zero for $x_{1}<0$. Asymmetric loading are applied to the crack faces (see Fig.(\ref{geom_frac})), thus the physical traction acting at the interface on the entire $x_{1}$ axis can be written as:
\beq
\BGs(x_1,x_2=0^+)=\Bp^+(x_{1})+\BGj(x_{1}), \quad \BGs(x_1,x_2=0^-)=\Bp^-(x_{1})+\BGj(x_{1}),
\eequ{tract_tot}

According to the approach illustrated in \citet{Wilmov1} and \citet{PiccMish2}, in order to derive explicit formulas for calculating the coefficients for the asymptotic of the stress fields near to the crack tip, we apply the Betti formula to the physical fields and to weight functions. The following relations, respectively in the upper and in the lower-midplane, are obtained:
\begin{eqnarray}
\int_{x_{2}=0^+} \left\{\BU^T(x_{1}^{'}-x_1,0^+)\BCR\BGs(x_1,0^+)-\BGS^T(x_{1}^{'}-x_1,0^+)\BCR\Bu(x_1,0^+)\right\}dx_1 & = & 0 \label{betti+}\\
\int_{x_{2}=0^-} \left\{\BU^T(x_{1}^{'}-x_1,0^-)\BCR\BGs(x_1,0^-)-\BGS^T(x_{1}^{'}-x_1,0^-)\BCR\Bu(x_1,0^-)\right\}dx_1 & = & 0 \label{betti-}
\end{eqnarray}
where $\BCR$ is the rotation matrix:
$$
\BCR=\pmatrix{-1 & 0 \cr 0 & 1},
$$
Subtracting the ($\ref{betti-}$) from the ($\ref{betti+}$) and using
%injecting the traction components along the $x_1$ axis 
\eq{tract_tot}, we derive  the following integral identity:
$$
\int_{x_{2}=0} \left\{\BU^T(x_{1}^{'}-x_1,0^+)\BCR\Bp^+(x_1)+\BU^T(x_{1}^{'}-x_1,0^+)\BCR\BGj(x_1)-\BU^T(x_{1}^{'}-x_1,0^-)\BCR\Bp^-(x_1)-\right.
$$
$$
\left.-\BU^T(x_{1}^{'}-x_1,0^-)\BCR\BGj(x_1)-[\BGS^T(x_{1}^{'}-x_1,0^+)\BCR\Bu(x_1,0^+)+\BGS^T(x_{1}^{'}-x_1,0^-)\BCR\Bu(x_1,0^-)]\right\}dx_1=0 
$$
%Rembering the continuity of the physical and of the singular tractions and the definition of displacemnte jump across the crack faces $[\Bu]$, and introducing the symmetric and skew-symmetric part of the loading \eq{p_symskew} and the weight functions (\ref{Usymd}) and (\ref{Uskewd}), we finally get:
Referring to  (\ref{Usymd}) and (\ref{Uskewd}) we deduce
$$
\int_{x_{2}=0} \left\{[\BU]^T(x_{1}^{'}-x_1,0^+)\BCR\BGj(x_1)-\BGS^T(x_{1}^{'}-x_1,0^+)\BCR[\Bu](x_1)\right\}=
$$
\beq
=-\int_{x_{2}=0} \left\{[\BU]^T(x_{1}^{'}-x_1,0^+)\BCR\left\langle \Bp\right\rangle(x_1)-\BGS^T(x_{1}^{'}-x_1,0^+)\BCR[\Bp](x_1)\right\}.
\eequ{Betti_real}

Let us define the Fourier transform of the skew-symmetric weight function and associated traction $\BGS$ respect to $x_{1}$:
\beq
[\hat{\BU}]^{+}(\xi)=\int^{+\infty}_{0}[\BU](x_{1})e^{i\xi x_{1}}\textrm{d}x_{1}, \quad \hat{\BGS}^{-}(\xi)=\int^{0}_{-\infty}\BGS(x_{1})e^{i\xi x_{1}}\textrm{d}x_{1},
\eequ{transf_sigfields}
Where the superscript $+$ indicates that $[\hat{\BU}]^{-}(\xi)$ is analytical in the upper half-plane ($\textrm{Im}\xi\in(0,+\infty)$) and the superscript $-$ indicates that $\hat{\BGS}^{-}(\xi)$ is analytical in the lower half-plane ($\textrm{Im}\xi\in(-\infty,0)$). The physical traction and the jump function are defined in such a way that the transforms $\hat{\BGj}^+(\xi)$ and $[\hat{\Bu}]^-(\xi)$ are analytic  in the upper and in the lower half-plane, respectively. Applying the Fourier transform to the \eq{Betti_real}, by means of convolutions properties, the following relation, valid for $\xi\in \textrm{R}$, is derived \citep{PiccMish3,PiccMish2}:
\beq
[\hat{\BU}]^{+T}\BCR\hat{\BGj}^{+}-\hat{\BGS}^{-T}\BCR[\hat{\Bu}]^{-}=-[\hat{\BU}]^{+T}\BCR\langle\hat{\Bp}\rangle-\langle\hat{\BU}\rangle^{+T}\BCR[\hat{\Bp}],\quad \xi\in \textrm{R},
\eequ{Betti}
This identity, obtained by \citet{Wilmov1} and \citet{PiccMish2}, relates transforms of the physical solutions to transforms of the weight functions, and it is used  for the evaluation of the stress identity factors.
\subsection{Asymptotic representation of the fields and stress intensity factors}
\label{asym_k}
%In references \cite{MishKuhn1,PiccMish4} and \cite{PiccMish3} it is explained how the dispacements and stress fields solutions of the interfacial plane crack problem in isotropic media with locally bounded energy can be represented in terms of Mellin trasforms ad how this representation leads to asymptotic expressions of the fields near the crack tip, where the stress become singular. Extending this approach to intercacial cracks problem between anisotropic materials, the 
The physical traction \eq{tract} and the displacement jump across the crack face \eq{jump} for $x_{1}\rightarrow 0$,  can be written in  the matrix form, as follows:
\begin{eqnarray}
\BGj(x_{1})             & = & \fr{x_{1}^{-\fr{1}{2}}}{2\sqrt{2\pi}}\BCT(x_1)\BK+\fr{x_{1}^{\fr{1}{2}}}{2\sqrt{2\pi}}\BCT(x_1)\BJ+\fr{x_{1}^{\fr{3}{2}}}{2\sqrt{2\pi}}\BCT(x_1)\BL+\CO(x_{1}^{\fr{5}{2}})\label{tract_pert}\\
\left[\Bu\right](x_{1}) & = & \fr{(-x_{1})^{\fr{1}{2}}}{\sqrt{2\pi}}\BCU(x_1)\BK+\fr{(-x_{1})^{\fr{3}{2}}}{\sqrt{2\pi}}\BCU(x_1)\BJ+\fr{(-x_{1})^{\fr{5}{2}}}{\sqrt{2\pi}}\BCU(x_1)\BL+\CO((-x_{1})^{\fr{7}{2}})\label{jump_pert}
\end{eqnarray}
Where $\BK=(K,\ov{K}),\BJ=(J,\ov{J}),\BL=(L,\ov{L})$, and $K$ and $L$ are higher order coefficients, defined in the same way of the stress intensity factor:  $J=J_I+iJ_{II}, L=L_I+iL_{II}$. Matrices $\BCT(x_1)$ and $\BCU(x_1)$ are given by:
$$
\BCT(x_1)=2\left(\Bw x_{1}^{i\Gve}, \ov{\Bw} x_{1}^{-i\Gve}\right),\quad \BCU(x_1)=\fr{2(\BH + \ov{\BH})}{\cosh \pi \Gve}\left(\fr{\Bw (-x_{1})^{i\Gve}}{1 + 2 i \Gve}, \fr{\ov{\Bw} (-x_{1})^{-i\Gve}}{1 - 2 i \Gve}\right),
$$

%From equations (\ref{tract_pert}) and (\ref{jump_pert}) we can note that stress intensity factor represents the coefficient of first order term of asymptotics of the physical solution of the problem. The evaluation of this factor and of higher order coefficient is an important issue for studying many different elastic cracks propagation problems. These calculations have been performed by 
The evaluation of coefficients in the asymptotic formulas  (\ref{tract_pert}) and (\ref{jump_pert}) is included in \citet{PiccMish3,PiccMish1} for interfacial cracks between dissimilar isotropic media, by means of a general integral formula involving  symmetric and skew-symmetric weight functions matrices. Here this approach is extended to anisotropic bi-materials: in the next Section general expressions for Fourier transforms of symmetric and skew-symmetric weight functions are derived using the Stroh formulation, as defined in Section (\ref{Stroh_int}), while in Section (\ref{ortho}) these expressions will be specialized to the case of a crack between two orthotropic materials \citep{Suo1} and the general formula obtained by Piccolroaz \emph{et al.} \citep{PiccMish3,PiccMish1} will be applied for calculating stress intensity factor for this case.

\section{Symmetric and skew-symmetric weight functions for anisotropic bi-materials}\label{weight_func}

Here we derive general expressions for symmetric and skew-symmetric weight function defined in Section (\ref{weight_betti})  as the jump $[\BU]$ and the average $\left\langle \BU\right\rangle$ of the singular solution for a semi-infinite interfacial crack with free-traction conditions.

Let us introduce the vector function $\Bh(z)$, solution of the Riemann-Hilbert problem for the physical traction \eq{eq5}, such that:
$$
\Bh(z) = \left\{ \begin{array}{cc}
\BB \Bg(z) , & \mbox{Im}~ z  \geq  0, \\
- \ov{\BB} \ov{\Bg}(z) , & \mbox{Im} ~ z < 0.
\end{array} \right.
$$
According to the Plemelj formula, $\Bh(z)$ can be written as follows:
$$
\Bh(z) = \fr{1}{2 \pi i}  \int_{-\infty}^\infty \fr{\BGj(\eta)}{\eta - z} d \eta,
$$
where the integral is understood in the principal value sense. Taking the Fourier transform respect to $x_{1}$, for $x_{2}=0^\pm$, at the interface, we obtain:
$$
\hat{\Bh}(\xi,x_{2}=0^\pm)=\hat{\Bh}^\pm (\xi) = \pm \int_{-\infty}^\infty e^{i \xi \eta} H(\pm \xi) \BGj(\eta) d \eta = \pm H(\mp \xi) \hat{\BGj}^+(\xi),\quad \xi\in\textrm{R},
$$
where $H$ is the Heaviside function, and the traction $\BGj$ is defined in such a way that its transform is analytic in the upper half-plane.
To derive the above representation we used the following relations
$$
\int_{-\infty}^\infty \fr{e^{i \xi (x_{1}\pm i0)}}{\eta - (x_{1} \pm i 0)} dx_{1} =
% \left[ \begin{array}{cc} 
%0 , & \xi > 0 \\
\pm 2 \pi i e^{i \xi \eta}  H(\mp \xi). %& \xi < 0.
%\end{array}  \right.
$$

Hence the Fourier transforms of $g$ and $\ov{g}$ at the interface are obtained in the form
$$
\hat{\Bg} (\xi)  = \BB^{-1} \hat{\Bh}^+ (\xi) = H(-\xi) \BB^{-1} \hat{\BGj}^+ (\xi), \quad \xi\in\textrm{R},
$$
and
$$
\hat{\ov{\Bg}} (\xi) = - \ov{\BB}^{-1} \hat{\Bh}^-(\xi) = H(\xi)  \ov{\BB}^{-1} \hat{\BGj}^+(\xi), \quad \xi\in\textrm{R}.
$$
By applying the Fourier transform to \eq{eq2} for the derivative of the physical displacements,  we deduce:
$$
- i \xi \hat{\Bu}(\xi,x_{2}=0^\pm)    = \BA \hat{\Bg}(\xi) + \ov{\BA} \hat{\ov{\Bg}}(\xi),\quad \xi\in\textrm{R}
$$
Hence, the Fourier transform of the displacements on the boundary of the upper half-plane is given by:
$$
\hat{\Bu}(\xi,x_{2}=0^+) = \fr{i}{\xi} \Big\{  H(-\xi) \BA \BB^{-1} + H(\xi) \ov{\BA}~ \ov{\BB}^{-1}\Big\}~ \hat{\BGj}^+ (\xi) 
$$
$$
= \fr{1}{\xi} \Big\{  H(-\xi) \BY^{(1)} - H(\xi) \ov{\BY}^{(1)}  \Big\} \hat{\BGj}^+(\xi) .
$$
By expressing the Heaviside function in the form:
$$
H(\pm\xi)  = \fr{1}{2} \Big( 1 \pm \mbox{sign} ~ \xi    \Big)
$$
we obtain:
\beq
\hat{\Bu} (\xi,x_{2}=0^+) = \Big\{  \fr{1}{2 \xi} (\BY^{(1)} - \ov{\BY}^{(1)}) - \fr{1}{2 |\xi |}     (  \BY^{(1)} + \ov{\BY}^{(1)}  )   \Big\} \hat{\BGj}^+(\xi),\quad \xi\in\textrm{R}.
\eequ{u+}
Following the same pattern of the derivation as above, we derive the Fourier transform of the physical displacement on the boundary of the lower half-plane:
\beq
\hat{\Bu} (\xi,x_{2}=0^-) = \Big\{  \fr{1}{2 \xi} (\BY^{(2)} - \ov{\BY}^{(2)}) + \fr{1}{2 |\xi |}     (  \BY^{(2)} + \ov{\BY}^{(2)}  )   \Big\} \hat{\BGj}^+(\xi),\quad \xi\in\textrm{R} . 
\eequ{u-}

Next, we replace $\Bu$ in the above text by the singular solution $\BU$ for a semi-infinite interfacial crack. Correspondingly, the vector of tractions $\BGj$ is replaced by $\BGS$, which is the traction vector corresponding to the singular solution $\BU$.  The Fourier transforms of the singular displacements on the boundary can then be derived:
\beq
\hat{\BU} (\xi,x_{2}=0^+) = \Big\{  \fr{1}{2 \xi} (\BY^{(1)} - \ov{\BY}^{(1)}) - \fr{1}{2 |\xi |}     (  \BY^{(1)} + \ov{\BY}^{(1)}  )   \Big\} \hat{\BGS}^-(\xi),\quad \xi\in\textrm{R}.
\eequ{U+}
\beq
\hat{\BU} (\xi,x_{2}=0^-) = \Big\{  \fr{1}{2 \xi} (\BY^{(2)} - \ov{\BY}^{(2)}) + \fr{1}{2 |\xi |}     (  \BY^{(2)} + \ov{\BY}^{(2)}  )   \Big\} \hat{\BGS}^-(\xi),\quad \xi\in\textrm{R} . 
\eequ{U-}
According to definitions (\ref{Usymd}) and (\ref{Uskewd}), we take the difference and the average of \eq{U+} and \eq{U-}, and we derive that:
\beq
[\hat{\BU}]^+(\xi)= %\hat{\Bu}^+  - \hat{\Bu}^-  = 
\fr{1}{|\xi|} \Big\{ i ~ \mbox{sign} (\xi) ~ \mbox{Im} (\BY^{(1)}- \BY^{(2)})  - \mbox{Re} (\BY^{(1)} + \BY^{(2)})   \Big\} \hat{\BGS}^-(\xi) ,
\eequ{eq6}
and
\beq
\langle \hat{\BU} \rangle(\xi) = \fr{1}{2 |  \xi |} \Big\{ i ~ \mbox{sign} (\xi) ~ \mbox{Im} (\BY^{(1)}+ \BY^{(2)})  - \mbox{Re} (\BY^{(1)} - \BY^{(2)})   \Big\} \hat{\BGS}^-(\xi),\quad \xi\in\textrm{R} .
\eequ{eq7}
We note that \eq{eq6} is the functional equation of the Wiener-Hopf type, similar to the one studied in \citet{PiccMish1,PiccMish2} and \citet{PiccMish3} for the case of isotropic media.  

Using the continuity of tractions $\BGS$ across  the interface, together with \eq{eq6} and \eq{eq7}, we express the Fourier transform of the skew-symmetric weight function $\langle \BU \rangle$ in the form:
\beq
\langle \hat{\BU} \rangle(\xi) = \BCA  [\hat{\BU}]^+(\xi)  + \fr{i}{\xi} \BCB \hat{\BGS}^-(\xi), \quad \xi\in\textrm{R},
\eequ{skew_wf}
where the diagonal matrix $\BCA$ and the off-diagonal matrix $\BCB$ are
$$
\BCA = \fr{1}{2} ~\mbox{Re}(\BY^{(1)} - \BY^{(2)})  \Big(  \mbox{Re}(\BY^{(1)} + \BY^{(2)}) \Big)^{-1},
$$
and
$$
\BCB = \fr{1}{2} ~\mbox{Im}(\BY^{(1)} + \BY^{(2)})   -  \BCA ~  \mbox{Im}(\BY^{(1)} - \BY^{(2)}). 
$$

The  formulas \eq{eq6} and  \eq{eq7}  give 
%(or alternatively \eq{skew_wf}) 
% general 
expressions for the Fourier transform of the symmetric and skew-symmetric weight functions for interfacial cracks in anisotropic bi-materials in terms of the transformed singular traction $\hat{\BGS}$. These expressions are compared to those obtained by the construction of the full-field singular solution for the elasticity problem in an half-plane \citep{PiccMish1} in Appendix A. 
%The perfect equality detected 
As expected, the weight functions derived via two different approaches agree.  Making the inversion of the \eq{eq6} and the \eq{eq7} the explicit representation of the weight functions can be obtained and the complex stress intensity factors associated to an arbitrary system of forces can be evaluated by means of integral formulas derived by the application of the Betty identity \citep{PiccMish1}.
\section{Interfacial crack in orthotropic bi-materials}\label{ortho}
In this section explicit expressions for the Fourier transforms of the symmetric and skew-symmetric weight functions for an interfacial crack in orthotropic bi-materials are derived using equations \eq{eq6} and \eq{eq7}. 
%These transforms, or alternatively the explicit representations derived by the inversion, can then be used to evaluate the complex stress identity factor $K$ by means of integral expressions derived from the Betti formula \cite{PiccMish1,PiccMish3}.
\subsection{Stroh representation for orthotropic bi-materials}
  For the case of orthotropic two-dimensional media, the matrices $\BQ, \BR$ and $\BT$ introduced in section (\ref{Stroh_int}) are given  by:
$$
\BQ = \pmatrix{c_{11} & 0 \cr 0 & c_{66}}, \BR = \pmatrix{0 & c_{12} \cr c_{66} & 0}, \BT = \pmatrix{c_{66} & 0 \cr 0 & c_{22}} .
$$ 
Where $c_{ij}$ are the elements of the materials stiffness matrix, which can be expressed in function of the elements of the compliance matrix $s_{ij}$ \citep{Suo1}:
\begin{eqnarray}
c_{11} & = & -\frac{s_{22}}{s_{12}^{2}-s_{11}s_{22}}\quad  c_{12}=\frac{s_{12}}{s_{12}^{2}-s_{11}s_{22}}\\
c_{22} & = & -\frac{s_{11}}{s_{12}^{2}-s_{11}s_{22}}\quad  c_{66}=\frac{1}{s_{66}}
\end{eqnarray}
 
 The characteristic equation \eq{eq5} becomes:
\beq
s_{11}\mu^{4}+(2s_{12}+s_{66})\mu^{2}+s_{22}=0
\eequ{char_eq} 
 The matrices $\BA$ and $\BB$, defined by relations \eq{eq4} and \eq{eq4}, are equivalent to those provided by alternative Lekhnitskii formulation \citep{Lekh1}, more precisely the 
Lekhnitskii approach gives a specially normalized eigenvector matrix $\BA$, and the characteristics equation derived using the Lekhnitskii formalism is identical to the \eq{char_eq} \citep{Suo1, Ting1}. The relationships between the two alternative formulations are derived and reported in details in reference \cite{Hwu1}, where the elements of  $\BA$ and $\BB$ in the Stroh representation are reported in function of the coefficient proposed by Lekhnitskii. Here we write the Stroh matrices $\BA$ and $\BB$ using this particular normalization, reported in \citet{Hwu1}:
\beq
\BA = \pmatrix{\fr{s_{11}\mu_{1}^{2}+s_{12}}{\sqrt{\fr{2}{\mu_{1}}(s_{22}-s_{11}\mu_{1}^{4})}}&  \fr{s_{11}\mu_{2}^{2}+s_{12}}{\sqrt{\fr{2}{\mu_{2}}(s_{22}-s_{11}\mu_{2}^{4})}} \cr \fr{s_{12}\mu_{1}+\fr{s_{22}}{\mu_{1}}}{\sqrt{\fr{2}{\mu_{1}}(s_{22}-s_{11}\mu_{1}^{4})}} & \fr{s_{12}\mu_{2}+\fr{s_{22}}{\mu_{2}}}{\sqrt{\fr{2}{\mu_{1}}(s_{22}-s_{11}\mu_{2}^{4})}}}
\eequ{Amatrix} 
\beq
\BB = \pmatrix{\fr{-\mu_{1}}{\sqrt{\fr{2}{\mu_{1}}(s_{22}-s_{11}\mu_{1}^{4})}}&  \fr{-\mu_{2}}{\sqrt{\fr{2}{\mu_{2}}(s_{22}-s_{11}\mu_{2}^{4})}} \cr \fr{1}{\sqrt{\fr{2}{\mu_{1}}(s_{22}-s_{11}\mu_{1}^{4})}} & \fr{1}{\sqrt{\fr{2}{\mu_{1}}(s_{22}-s_{11}\mu_{2}^{4})}}}
\eequ{Bmatrix} 
Where $\mu_{1}$ and $\mu_{2}$ are the two roots of the characteristics equation with positive imaginary part. The Hermitian matrix $\BY$ evaluated using \eq{Amatrix} and \eq{Bmatrix} is:
\beq
\BY = i \BA \BB^{-1}= \pmatrix{s_{11}\textrm{Im}(\mu_{1}+\mu_{2}) & -i(s_{11}\mu_{1}\mu_{2}-s_{12}) \cr i(s_{11}\overline{\mu}_{1}\overline{\mu}_{2}-s_{12})  & -s_{22}\textrm{Im}\left(\fr{1}{\mu_{1}}+\fr{1}{\mu_{2}}\right)}
\eequ{Ymatrix}

Following the notation introduced by Suo, \citep{Suo1, GupArg1}, we define two adimentional parameters measuring the material anisotropy:
$$
\lambda=\fr{s_{11}}{s_{22}}\qquad \rho=\fr{1}{2}\fr{2s_{12}+s_{66}}{\sqrt{s_{11}s_{22}}}
$$

If $\lambda=1$ the material has transversely cubic symmetry, while if $\lambda=\rho=1$ the material is transversely isotropic. The positive definetess of the strain energy density requires that:
$$
\lambda>0, \qquad -1<\rho<+\infty,
$$
The characteristics equation \eq{char_eq} then becomes:
\beq
\lambda\mu^{4}+2\rho\lambda^{\fr{1}{2}}\mu^{2}+1=0
\eequ{char_eqsim}
The roots with positive imaginary parts are:
\begin{eqnarray}
\mu_{1}&=&i\lambda^{-\fr{1}{4}}(n+m),\quad \mu_{2}=i\lambda^{-\fr{1}{4}}(n-m),\quad\textrm{for}\quad 1 <\rho<+\infty,\nonumber\\
\mu_{1}&=&\lambda^{-\fr{1}{4}}(in+m),\quad \mu_{2}=\lambda^{-\fr{1}{4}}(in-m),\quad\textrm{for}\quad-1 <\rho<1,\nonumber\\
\mu_{1}&=&\mu_{2}=i\lambda^{-\fr{1}{4}},\qquad \qquad \qquad  \qquad  \qquad \quad  \textrm{for}\quad \rho=1,\nonumber
\end{eqnarray}
Where:
$$
n=\left(\fr{1}{2}(1+\rho)\right)^{\fr{1}{2}},\qquad m=\left|\fr{1}{2}(1-\rho)\right|^{\fr{1}{2}},
$$
Using this notation the matrix $\BY$ becomes:
\beq
\BY = i \BA \BB^{-1}= \pmatrix{2n\lambda^{\fr{1}{4}}(s_{11}s_{22})^{\fr{1}{2}} & i((s_{11}s_{22})^{\fr{1}{2}}+s_{12}) \cr -i((s_{11}s_{22})^{\fr{1}{2}}+s_{12})  & 2n\lambda^{-\fr{1}{4}}(s_{11}s_{22})^{\fr{1}{2}}}
\eequ{YmatrixSuo}
The bi-material matrix $\BH$ for two orthotropic materials \citep{Suo1, GupArg1} is:
\beq
\BH = \BY^{(1)}+\overline{\BY}^{(2)}= \pmatrix{H_{11} & -i\beta\sqrt{H_{11}H_{22}} \cr i\beta\sqrt{H_{11}H_{22}}  & H_{22}}
\eequ{HmatrixSuo}
Where:
\begin{eqnarray}
H_{11}                  &=& [2n\lambda^{\fr{1}{4}}(s_{11}s_{22})^{\fr{1}{2}}]^{(1)}+[2n\lambda^{\fr{1}{4}}(s_{11}s_{22})^{\fr{1}{2}}]^{(2)},\nonumber\\
H_{22}                  &=& [2n\lambda^{-\fr{1}{4}}(s_{11}s_{22})^{\fr{1}{2}}]^{(1)}+[2n\lambda^{-\fr{1}{4}}(s_{11}s_{22})^{\fr{1}{2}}]^{(2)},\nonumber\\         
\beta\sqrt{H_{11}H_{22}}&=& [((s_{11}s_{22})^{\fr{1}{2}}+s_{12})]^{(2)}-[((s_{11}s_{22})^{\fr{1}{2}}+s_{12})]^{(1)},\nonumber
\end{eqnarray}
$\beta$ is the generalization of one of the Dundurs parameters \citep{Dund1}, and is connected to the bi-material oscillatory index $\varepsilon$ by the relation:
$$
\varepsilon=\fr{1}{2\pi}\ln\left(\fr{1-\beta}{1+\beta}\right)
$$ 
The eigenvector $\Bw$ of the eigensystem \eq{eigens}, associated with the component of displacement jump \eq{jump} and of the traction ahead of the crack \eq{tract}, is assumed in the normalized form:
\beq
\Bw=\left(-\fr{1}{2}i,\fr{1}{2}\sqrt{\fr{H_{11}}{H_{22}}}\right)^{T},
\eequ{wvector}
\subsection{Weight functions for in-plane deformations}
In order to derive explicit expressions for the Fourier transforms of the weight functions matrices from the relations \eq{eq6} and \eq{eq7}, we need to evaluate the Fourier transform of the singular traction $\BGS$. The singular traction on the boundary $\BGS$ for an interfacial crack between two generic anisotropic materials is given by \eq{sigma}. Here we report this expression: 
\beq
\BGS=\fr{(-x_{1})^{-\fr{3}{2}}}{\sqrt{2\pi}}\textrm{Re}\left(C\ov{\Bw}(-x_{1})^{i\varepsilon}\right)
\eequ{sigma_rep}

 For studying a crack between two orthotropic media, we assume for $\Bw$ the expression \eq{wvector}. As just anticipated, mode I and mode II are coupled, and associated to one single complex constant $C$.  Therefore the spaces of singular tractions and of singular solutions $\BU$ are two-dimensional linear spaces, and two linearly independent vectors $\BGS$ and weight functions must be defined \citep{PiccMish1,PiccMish3}. Assuming $C_{I}=1,C_{II}=0$, and $C_{I}=0,C_{II}=1$, and substituing explicit components of $\Bw$, the two independent traction vectors are defined:
\beq
\BGS^{1}(x_{1})= \fr{(-x_{1})^{-\fr{3}{2}}}{2\sqrt{2\pi}}\pmatrix{i((-x_{1})^{i\Gve}-(-x_{1})^{-i\Gve}) \cr\cr \sqrt{\fr{H_{11}}{H_{22}}}((-x_{1})^{i\Gve}+(-x_{1})^{-i\Gve})  };
\eequ{trac_sing1}
\beq
\BGS^{2}(x_{1})= \fr{(-x_{1})^{-\fr{3}{2}}}{2\sqrt{2\pi}}\pmatrix{-((-x_{1})^{i\Gve}+(-x_{1})^{-i\Gve}) \cr\cr i\sqrt{\fr{H_{11}}{H_{22}}}((-x_{1})^{i\Gve}-(-x_{1})^{-i\Gve})  };
\eequ{trac_sing2}
Applying the Fourier transform to \eq{trac_sing1} and \eq{trac_sing2}, the result is:
\beq
\hat{\BGS}^{1-}(\xi)= \fr{\xi_{-}^{\fr{1}{2}}}{1+4\Gve^{2}}\pmatrix{-\fr{e_{0}\xi_{-}^{-i\Gve}}{c^{+}}\left(-\fr{1}{2}-i\Gve\right)+\fr{\xi_{-}^{i\Gve}}{e_{0}c^{-}}\left(-\fr{1}{2}+i\Gve\right)\cr\cr i\sqrt{\fr{H_{11}}{H_{22}}}\left\{\fr{e_{0}\xi_{-}^{-i\Gve}}{c^{+}}\left(-\fr{1}{2}-i\Gve\right)+\fr{\xi_{-}^{i\Gve}}{e_{0}c^{-}}\left(-\fr{1}{2}+i\Gve\right)\right\}  };
\eequ{transf_sigma1}
\beq
\hat{\BGS}^{2-}(\xi)= \fr{\xi_{-}^{\fr{1}{2}}}{1+4\Gve^{2}}\pmatrix{-i\left\{\fr{e_{0}\xi_{-}^{-i\Gve}}{c^{+}}\left(-\fr{1}{2}-i\Gve\right)+\fr{\xi_{-}^{i\Gve}}{e_{0}c^{-}}\left(-\fr{1}{2}+i\Gve\right)\right\}\cr\cr \sqrt{\fr{H_{11}}{H_{22}}}\left\{-\fr{e_{0}\xi_{-}^{-i\Gve}}{c^{+}}\left(-\fr{1}{2}-i\Gve\right)+\fr{\xi_{-}^{i\Gve}}{e_{0}c^{-}}\left(-\fr{1}{2}+i\Gve\right)\right\} };
\eequ{transf_sigma2}
Where $\textrm{Im}\xi_{-}\in(-\infty,0)$, $e_{0}=e^{\Gve\fr{\pi}{2}}$ and:
$$
e_{0}=e^{\Gve\fr{\pi}{2}}, \quad c^{\pm}=\fr{(1+i)\sqrt{\pi}}{2\Gamma\left[\fr{1}{2}\pm i\Gve\right]},
$$
Substituing the  \eq{transf_sigma1} and \eq{transf_sigma1} into \eq{eq6} and \eq{eq7}, and expressing the matrices in term of elements of $\BH$, we derive the Fourier transforms of the two independent symmetric and skew symmetrix weight functions:
\begin{equation}
\pmatrix{[\hat{U}^{1}_{1}]^{+} & [\hat{U}^{2}_{1}]^{+} \cr\cr [\hat{U}^{1}_{2}]^{+} & [\hat{U}^{2}_{2}]^{+}}=-\fr{\sqrt{H_{11}H_{22}}}{|\xi|}\pmatrix{\sqrt{\fr{H_{11}}{H_{22}}} & i\beta\mbox{sign}(\xi) \cr\cr -i\beta\mbox{sign} (\xi)  & \sqrt{\fr{H_{22}}{H_{11}}}} \pmatrix{[\hat{\GS}^{1}_{1}]^{-} & [\hat{\GS}^{2}_{1}]^{-} \cr\cr [\hat{\GS}^{1}_{2}]^{-} & [\hat{\GS}^{2}_{2}]^{-}};
\label{symm_wf}
\end{equation}
\begin{equation}
\pmatrix{\langle\hat{U}^{1}_{1}\rangle & \langle\hat{U}^{2}_{1}\rangle \cr\cr \langle\hat{U}^{1}_{2}\rangle & \langle\hat{U}^{2}_{2}\rangle}=-\fr{\sqrt{H_{11}H_{22}}}{2|\xi|}\pmatrix{\delta_{1}\sqrt{\fr{H_{11}}{H_{22}}} & -i\gamma\mbox{sign}(\xi) \cr\cr i\gamma\mbox{sign} (\xi)  & \delta_{2}\sqrt{\fr{H_{22}}{H_{11}}}} \pmatrix{[\hat{\GS}^{1}_{1}]^{-} & [\hat{\GS}^{2}_{1}]^{-} \cr\cr [\hat{\GS}^{1}_{2}]^{-} & [\hat{\GS}^{2}_{2}]^{-}};
\label{skew_wf}
\end{equation}
Where the following Dundurs-like parameters have been defined:
\begin{eqnarray}
\delta_{1}                  &=& \fr{[2n\lambda^{\fr{1}{4}}(s_{11}s_{22})^{\fr{1}{2}}]^{(1)}-[2n\lambda^{\fr{1}{4}}(s_{11}s_{22})^{\fr{1}{2}}]^{(2)}}{H_{11}},\nonumber\\
\delta_{2}                  &=& \fr{[2n\lambda^{-\fr{1}{4}}(s_{11}s_{22})^{\fr{1}{2}}]^{(1)}-[2n\lambda^{-\fr{1}{4}}(s_{11}s_{22})^{\fr{1}{2}}]^{(2)}}{H_{22}},\nonumber\\      
\gamma                      &=& \fr{[((s_{11}s_{22})^{\fr{1}{2}}+s_{12})]^{(1)}+[((s_{11}s_{22})^{\fr{1}{2}}+s_{12})]^{(2)}}{\sqrt{H_{11}H_{22}}},\nonumber
\end{eqnarray}
 The superscript $+$ indicates that $\hat{\BU}^{+}(\xi)$ is analytic in the upper half-plane ($\textrm{Im}\xi\in(0,+\infty)$). The expression (\ref{symm_wf}) for the transform of the symmetric weight functions, as just anticipated in previous session, is a Wiener-Hopf type equation similar to the one solved for interfacial crack in isotropic materials in \citet{PiccMish3,PiccMish2,PiccMish1}. The skew-symmetric part of the weight function can also be decomposed by means of relation \eq{skew_wf}:
\beq
\langle \hat{\BU} \rangle = \BCA  [\hat{\BU}]^{+}  + \fr{i}{\xi} \BCB \hat{\BGS}^{-},
\eequ{skew_wfRH}
Where the matrices $\BCA$ and $\BCB$ for orthotropic bi-materials using our notation are given by:
$$
\BCA=\fr{1}{2}\pmatrix{\delta_{1} & 0 \cr 0 & \delta_{2}}; \quad \BCB=\fr{\sqrt{H_{11}H_{22}}}{2}\pmatrix{0 & \gamma+\beta\delta_{1} \cr -(\gamma+\beta\delta_{2}) & 0};
$$
The explicit expressions for the weight functions are obtained by the inversion of the derived transforms. The symmetric weight function $[\BU](x_{1})$ is equal to zero for $x_{1}<0$, while for $x_{1}>0$ it is given by:
\begin{eqnarray}
\left[U^{1}_{1}\right](x_{1}) & = & \fr{H_{11}x_{1}^{-\fr{1}{2}}}{2c^{+}c^{-}\sqrt{2\pi}(1+4\Gve^{2})}\left\{(\beta-1)\left(-\fr{1}{2}+i\Gve\right)x_{1}^{-i\Gve}+(\beta+1)\left(-\fr{1}{2}-i\Gve\right)x_{1}^{i\Gve}\right\}\nonumber\\
\left[U^{1}_{2}\right](x_{1}) & = & \fr{i\sqrt{H_{11}H_{22}}x_{1}^{-\fr{1}{2}}}{2c^{+}c^{-}\sqrt{2\pi}(1+4\Gve^{2})}\left\{(\beta-1)\left(-\fr{1}{2}+i\Gve\right)x_{1}^{-i\Gve}-(\beta+1)\left(-\fr{1}{2}-i\Gve\right)x_{1}^{i\Gve}\right\}\nonumber\\
\left[U^{2}_{1}\right](x_{1}) & = & -\fr{iH_{11}x_{1}^{-\fr{1}{2}}}{2c^{+}c^{-}\sqrt{2\pi}(1+4\Gve^{2})}\left\{(\beta-1)\left(-\fr{1}{2}+i\Gve\right)x_{1}^{-i\Gve}-(\beta+1)\left(-\fr{1}{2}-i\Gve\right)x_{1}^{i\Gve}\right\}\nonumber\\
\left[U^{2}_{2}\right](x_{1}) & = & \fr{\sqrt{H_{11}H_{22}}x_{1}^{-\fr{1}{2}}}{2c^{+}c^{-}\sqrt{2\pi}(1+4\Gve^{2})}\left\{(\beta-1)\left(-\fr{1}{2}+i\Gve\right)x_{1}^{-i\Gve}+(\beta+1)\left(-\fr{1}{2}-i\Gve\right)x_{1}^{i\Gve}\right\}\nonumber\\
\end{eqnarray}
The skew-symmetric weight function $\langle\BU\rangle(x_{1})$, is equal to $\BCA[\BU](x) $  for $x_{1}>0$, while for $x_{1}<0$ it is given by:
\begin{eqnarray}
\langle U^{1}_{1}\rangle(x_{1}) & = & -\fr{iH_{11}(\gamma+\beta\delta_{1})(-x_{1})^{-\fr{1}{2}}}{4c^{+}c^{-}\sqrt{2\pi}(1+4\Gve^{2})}\left\{\left(-\fr{1}{2}+i\Gve\right)\fr{(-x_{1})^{-i\Gve}}{e_{0}^{2}}+\left(-\fr{1}{2}-i\Gve\right)e_{0}^{2}(-x_{1})^{i\Gve}\right\}\nonumber\\
\langle U^{1}_{2}\rangle(x_{1}) & = & \fr{\sqrt{H_{11}H_{22}}(\gamma+\beta\delta_{2})(-x_{1})^{-\fr{1}{2}}}{4c^{+}c^{-}\sqrt{2\pi}(1+4\Gve^{2})}\left\{ \left(-\fr{1}{2}+i\Gve\right)\fr{(-x_{1})^{-i\Gve}}{e_{0}^{2}}-\left(-\fr{1}{2}-i\Gve\right)e_{0}^{2}(-x_{1})^{i\Gve}\right\}\nonumber\\
\langle U^{2}_{1}\rangle(x_{1}) & = & -\fr{H_{11}(\gamma+\beta\delta_{1})(-x_{1})^{-\fr{1}{2}}}{4c^{+}c^{-}\sqrt{2\pi}(1+4\Gve^{2})}\left\{\left(-\fr{1}{2}+i\Gve\right)\fr{(-x_{1})^{-i\Gve}}{e_{0}^{2}}-\left(-\fr{1}{2}-i\Gve\right)e_{0}^{2}(-x_{1})^{i\Gve}\right\}\nonumber\\
\langle U^{2}_{2}\rangle(x_{1}) & = & -\fr{i\sqrt{H_{11}H_{22}}(\gamma+\beta\delta_{2})(-x_{1})^{-\fr{1}{2}}}{4c^{+}c^{-}\sqrt{2\pi}(1+4\Gve^{2})}\left\{\left(-\fr{1}{2}+i\Gve\right)\fr{(-x_{1})^{-i\Gve}}{e_{0}^{2}}+\left(-\fr{1}{2}-i\Gve\right)e_{0}^{2}(-x_{1})^{i\Gve}\right\}\nonumber\\
\end{eqnarray}
\subsection{Stress-intensity factor for orthotropic bi-materials}
The symmetric and skew-symmetric weight functions are now used for evaluating the complex stress factor for a loading correspondent to an arbitrary system of forces, following the procedure outlined in \citet{PiccMish2,PiccMish1} and \citet{PiccMish4}. In Section \ref{asym_k}, we have shown that complex stress identity factor represents the coefficient of the first order asymptotic term of physical traction field near the crack tip. Remembering equation (\ref{tract_pert}), for $x_2\rightarrow 0^+$ the traction becomes:
\beq
\BGj(x_{1})=\fr{x_{1}^{-\fr{1}{2}}}{2\sqrt{2\pi}}\BCT(x_1)\BK+\fr{x_{1}^{\fr{1}{2}}}{2\sqrt{2\pi}}\BCT(x_1)\BJ+\fr{x_{1}^{\fr{3}{2}}}{2\sqrt{2\pi}}\BCT(x_1)\BL+\CO(x_{1}^{\fr{5}{2}})
\eequ{tract+}
For orthotropic bi-materials, considering the components of $\Bw$ given by \eq{wvector}, the matrix $\BCT(x_{1})$ is:
$$
\BCT(x_{1})=\pmatrix{-ix_{1}^{i\Gve} & ix_{1}^{-i\Gve} \cr \cr \sqrt{\fr{H_{11}}{H_{22}}}x_{1}^{i\Gve} & \sqrt{\fr{H_{11}}{H_{22}}}x_{1}^{-i\Gve}},
$$
The Fourier transform of the \eq{tract+}, as $\xi\rightarrow\infty$ and $\textrm{Im}\xi\in(0,+\infty)$, is:
\beq
\hat{\BGj}^{+}(\xi)=\fr{\xi_{+}^{-\fr{1}{2}}}{4}\hat{\BCT}(\xi_{+})\BK+\fr{\xi_{+}^{-\fr{1}{2}}}{4\xi}\hat{\BCT}(\xi_{+})\BJ+\fr{\xi_{+}^{-\fr{1}{2}}}{4\xi^2}\hat{\BCT}(\xi_{+})\BL+\CO(\xi^{-\fr{7}{2}})
\eequ{tract_fourier}
Where:
$$
\hat{\BCT}(\xi_{+})=\pmatrix{\fr{\xi_{+}^{-i\Gve}}{c^{+}e_{0}} & -\fr{e_{0}\xi_{+}^{i\Gve}}{c^{-}} \cr \cr \sqrt{\fr{H_{11}}{H_{22}}}\fr{i\xi_{+}^{-i\Gve}}{c^{+}e_{0}} & \sqrt{\fr{H_{11}}{H_{22}}}\fr{ie_{0}\xi_{+}^{i\Gve}}{c^{-}}}
$$

Using this expression, the explicit transforms of the traction $\hat{\BGS}^{-}$ and of the symmetric weight functions matrix $[\hat{\BU}]^{+}$ derived in previous Section, and evaluating $[\hat{\Bu}]^{-}$ ,
the asymptotic expansions for the members of Betti identity \eq{Betti} are derived:
\begin{eqnarray}
[\hat{\BU}]^{+T}\BCR\hat{\BGj}^{+} & = & \xi^{-1}\BCM_{1}\BK+\xi^{-2}\BCM_{2}\BJ+\CO(\xi^{-3})\quad \textrm{for} \quad \textrm{Im}\xi\in(0,+\infty) \label{term1} \\
\hat{\BGS}^{-T}\BCR[\hat{\Bu}]^{-} & = & \xi^{-1}\BCM_{1}\BK+\xi^{-2}\BCM_{2}\BJ+\CO(\xi^{-3})\quad \textrm{for} \quad \textrm{Im}\xi\in(-\infty,0) \label{term2}
\end{eqnarray}
The explicit form for the matrix $\BCM_{1}$ is:
\beq
\BCM_{1}=-\fr{H_{11}}{4c^{+}c^{-}(1+4\Gve^{4})}\pmatrix{-\fr{(\beta-1)(1-2i\Gve)}{e_{0}^{2}} & e_{0}^{2}(\beta+1)(1+2i\Gve) \cr\cr \fr{i(\beta-1)(1-2i\Gve)}{e_{0}^{2}} &  ie_{0}^{2}(\beta+1)(1+2i\Gve)}
\eequ{Matrix}

Now, we rewrite the Fourier transform of the Betti identity \eq{Betti} in terms of a Riemann-Hilbert problem:
\beq
\mathbf{\Psi}^{+}(\xi)-\mathbf{\Psi}^{-}(\xi)=-[\hat{\BU}]^{+T}(\xi)\BCR\langle\hat{\Bp}\rangle(\xi)-\langle\hat{\BU}\rangle^{+T}(\xi)\BCR[\hat{\Bp}](\xi),\quad \xi\in\textrm{R}
\eequ{Betti_RH}
Where $\mathbf{\Psi}(\xi)$, according to the Plemelj formula, is:
\beq
\mathbf{\Psi}^{\pm}(\xi) = \fr{1}{2 \pi i}  \int_{-\infty}^\infty \fr{\mathbf{\Psi}(\eta)}{\eta - \xi} d \eta,
\eequ{Plemelj}
Then the solution of the \eq{Betti_RH} is:
$$
[\hat{\BU}]^{+T}\BCR\hat{\BGj}^{+}=\mathbf{\Psi}^{+},\quad \textrm{Im}\xi\in(0,+\infty);
$$
$$
\hat{\BGS}^{-T}\BCR[\hat{\Bu}]^{-} =\mathbf{\Psi}^{-},\quad \textrm{Im}\xi\in(-\infty,0);
$$
From these expressions asymptotic estimates can be extracted. For $\xi\rightarrow\infty$ we can expand Plemelj's formula, considering only the first term we have:
\beq
\mathbf{\Psi}^{\pm}(\xi) = \fr{1}{2 \pi i}  \int_{-\infty}^\infty \fr{\mathbf{\Psi}(\eta)}{\eta - \xi} d \eta=-\xi^{-1}\fr{1}{2 \pi i}  \int_{-\infty}^\infty \mathbf{\Psi}(\eta) d \eta+\CO(\xi^{-2})
\eequ{Plemelj_pert}
Substituing this expression and the expansions (\ref{term1}) and (\ref{term2}) in \eq{Betti_RH}, comparing the corresponding terms, and considering only the first order, the following general formula for the complex stress intensity factor is obtained:
\beq
\BK= \fr{1}{2 \pi i} \BCM_{1}^{-1} \int_{-\infty}^\infty \left\{ [\hat{\BU}]^{+T}(\eta)\BCR\langle\hat{\Bp}\rangle(\eta)+\langle\hat{\BU}\rangle^{+T}(\eta)\BCR[\hat{\Bp}](\eta) \right\}d \eta
\eequ{Stress_fact}
Since we have explicit expressions for matrix $\BCM_{1}$ and for the Fourier transforms of both symmetric and skew-symmetric weight functions, now using the formula \eq{Stress_fact} we can evaluate the complex stress intensity factor associated to an arbitrary loading for interfacial cracks in orthotropic bi-materials. In the next section an illustrative example of computation of $\BK$ by means of the  \eq{Stress_fact} is reported.
\section{An illustrative example}
\label{example}
\begin{figure}[htbp]
\centering
\includegraphics[width=5.5cm,height=6cm]{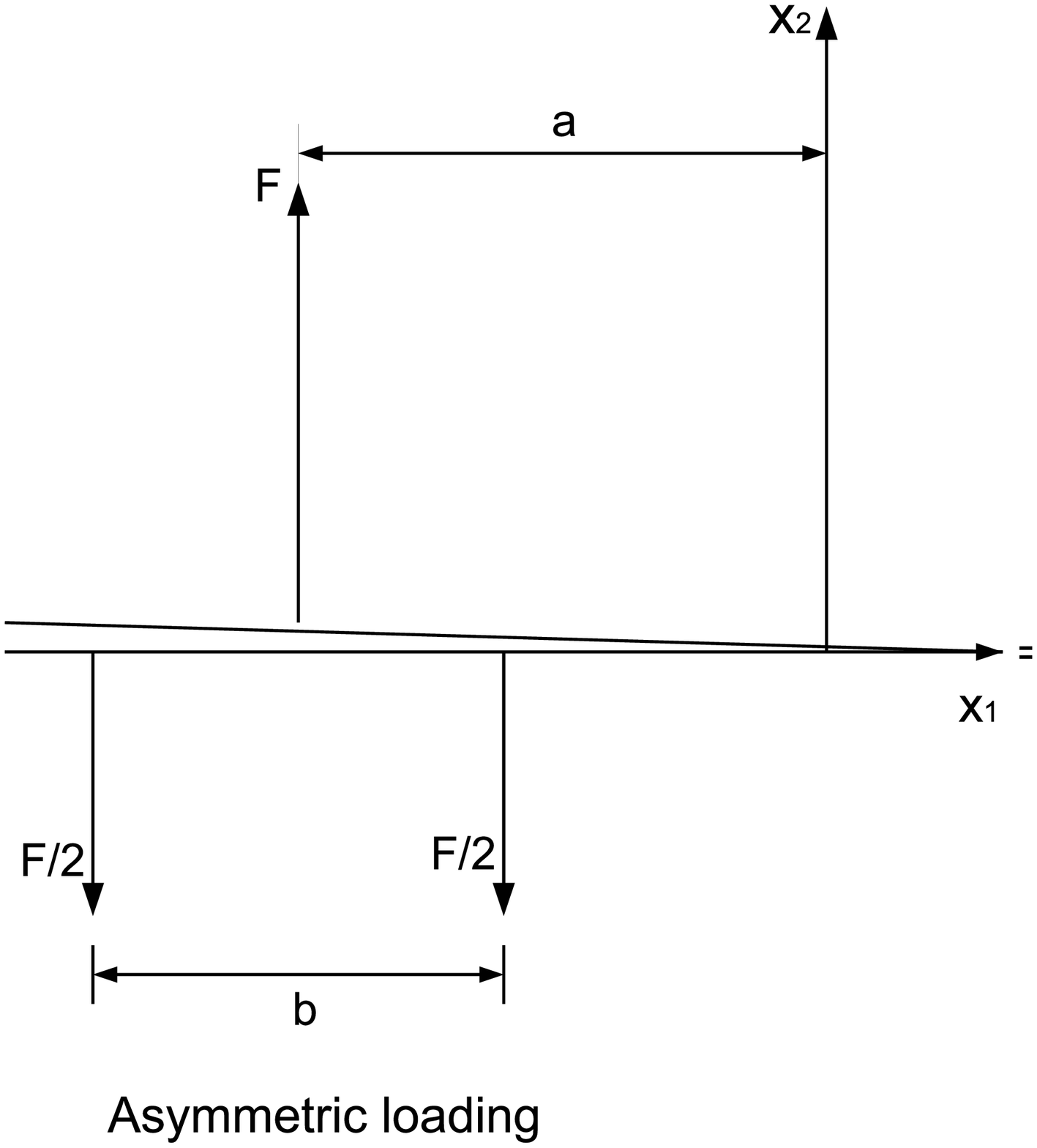}
\includegraphics[width=5.5cm,height=6cm]{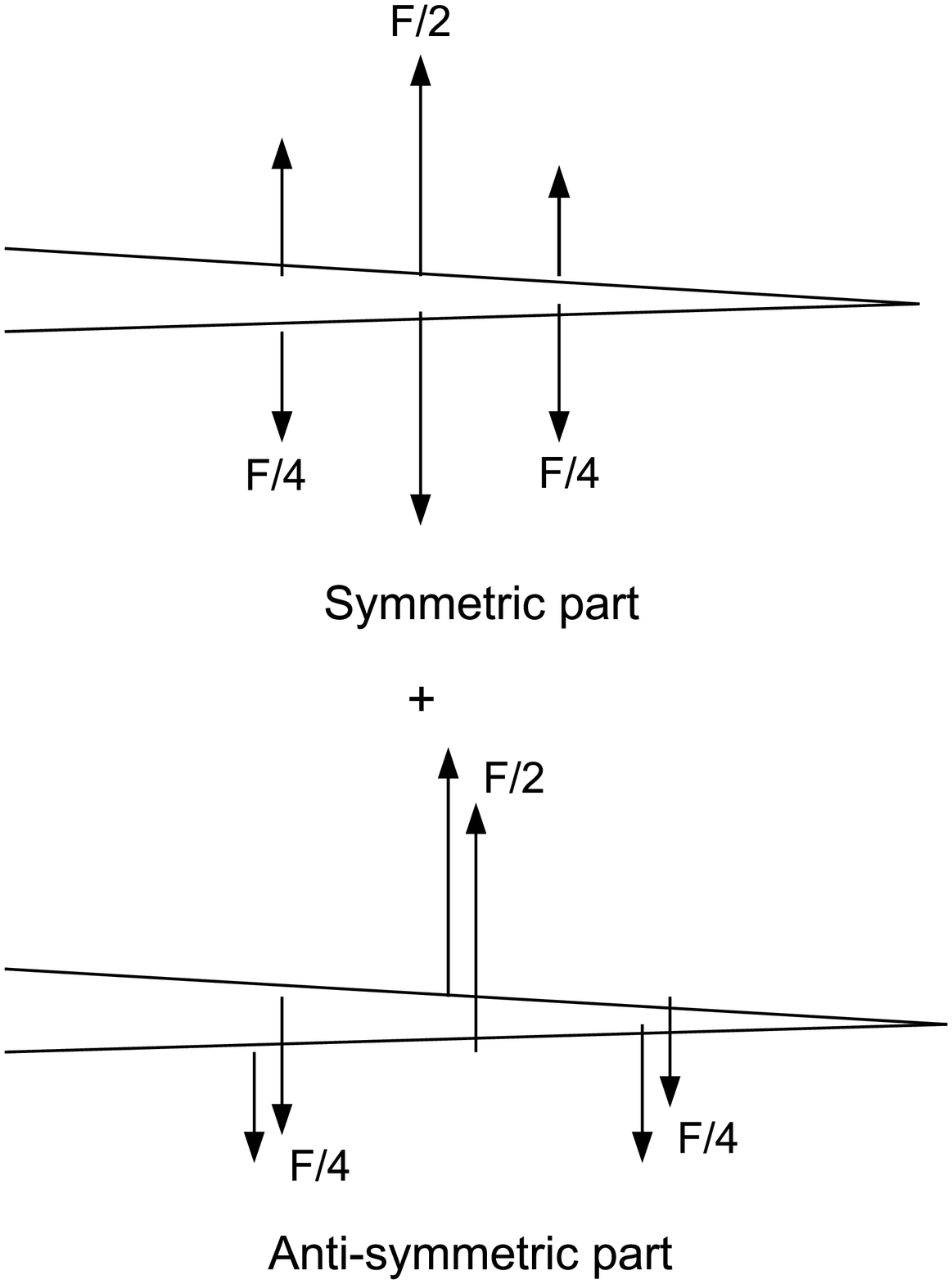}
\caption[loading]{\footnotesize {''Three point'' loading of an interfacial crack.}}
\label{loading}
\end{figure}

In this section we present an illustrative example of computation of complex stress intensity factor $K$ for an interfacial crack loaded by a simple asymmetric force system in orthotropic bi-materials. The considered force system is illustrated in Fig.(\ref{loading}): the loading consists in a point force $F$ acting upon the upper crack face at a distance $a$ behind the crack tip and two point forces $F/2$ acting upon the lower crack face at a distance $a-b/2$ and $a+b/2$, respectively, behind the crack tip. 

The loading can be expressed in terms of the Dirac delta function \citep{PiccMish1}:
\beq
p^{+}(x_{1})=-F\delta(x_{1}+a),\qquad p^{-}(x_{1})=-\fr{F}{2}\delta(x_{1}+a+b)-\fr{F}{2}\delta(x_{1}+a-b),
\eequ{load+-}
As it is shown in Fig.(\ref{loading}), the loading can be decomposed into symmetric and skew-symmetric part:
\begin{eqnarray}
\langle p\rangle(x_{1}) & = & -\fr{F}{2}\delta(x_{1}+a)-\fr{F}{4}\delta(x_{1}+a+b)-\fr{F}{4}\delta(x_{1}+a-b)      \nonumber\\
\left[p\right](x_{1})   & = & -F\delta(x_{1}+a)+\fr{F}{2}\delta(x_{1}+a+b)+\fr{F}{2}\delta(x_{1}+a-b)
\end{eqnarray}
The Fourier transform of the symmetric and skew symmetric part of the loading are given by:
\begin{eqnarray}
\langle\hat{p}\rangle(\xi) & = & -\fr{F}{2}e^{-i\xi a}-\fr{F}{4}e^{-i\xi(a+b)}-\fr{F}{4}e^{-i\xi(a-b)}   \nonumber\\
\left[\hat{p}\right](\xi)  & = & -Fe^{-i\xi a}+\fr{F}{2}e^{-i\xi(a+b)}+\fr{F}{2}e^{-i\xi(a-b)} 
\end{eqnarray}
Using these expressions and the explicit transforms of the symmetric and skew-symmetric weight functions in orthotropic media, (\ref{symm_wf}) and \eq{skew_wfRH}, both symmetric and anti-symmetric part of the complex stress intensity factor $K=K^{S}+K^{A}$ corresponding to this loading system have been evaluated by means of the integral formula \eq{Stress_fact}:
\begin{eqnarray}
K^{S} & = & \fr{e_{0}^{2}}{1-\beta}F\sqrt{\fr{H_{22}}{H_{11}}}\sqrt{\fr{2}{\pi}}a^{-\fr{1}{2}-i\Gve}\left\{\fr{1}{2}+\fr{1}{4}(1-b/a)^{-\fr{1}{2}-i\Gve}+\fr{1}{4}(1+b/a)^{-\fr{1}{2}-i\Gve}\right\}\nonumber\\
K^{A} & = & \fr{e_{0}^{2}(\gamma+\delta_{2})}{(1-\beta)^{2}}F\sqrt{\fr{H_{22}}{H_{11}}}\sqrt{\fr{2}{\pi}}a^{-\fr{1}{2}-i\Gve}\left\{\fr{1}{2}-\fr{1}{4}(1-b/a)^{-\fr{1}{2}-i\Gve}-\fr{1}{4}(1+b/a)^{-\fr{1}{2}-i\Gve}\right\}\nonumber
\end{eqnarray}
In order to study the behavior of these symmetric and anti-symmetric contributions to the stress intensity factors in function of $b/a$, the following non-dimensional parameters have been defined \citep{Suo1}:
$$
\GF=\fr{[(s_{11}s_{22})^{\fr{1}{2}}]^{(2)}}{[(s_{11}s_{22})^{\fr{1}{2}}]^{(1)}}, \GT^{(1)}=\left[\fr{s_{12}}{(s_{11}s_{22})^{\fr{1}{2}}}\right]^{(1)}, \GT^{(2)}=\left[\fr{s_{12}}{(s_{11}s_{22})^{\fr{1}{2}}}\right]^{(2)},
$$
Then we can exprime $H_{11},H_{22},\delta_{2}$ and $\gamma$ in function of these parameters:
$$
\fr{H_{22}}{H_{11}}=\fr{[2n\lambda^{-\fr{1}{4}}]^{(1)}+[2n\lambda^{-\fr{1}{4}}]^{(2)}\GF}{[2n\lambda^{\fr{1}{4}}]^{(1)}+[2n\lambda^{\fr{1}{4}}]^{(2)}\GF}
$$
$$
\delta_{2}=\fr{[2n\lambda^{-\fr{1}{4}}]^{(1)}-[2n\lambda^{-\fr{1}{4}}]^{(2)}\GF}{[2n\lambda^{-\fr{1}{4}}]^{(1)}+[2n\lambda^{-\fr{1}{4}}]^{(2)}\GF}
$$
$$
\gamma=\beta\fr{[1+\GT]^{(1)}-[1+\GT]^{(2)}\GF}{\GF[1+\GT]^{(1)}-[1+\GT]^{(2)}}
$$
It is important to note that the Dundurs parameter $\gamma$, associated to the skew-symmetric part of the loading, depends on $\beta$. Moreover, oscillations of the stress and displacement fields are excluded for $\beta=0$, $\Gve=0$ \citep{Suo1}. In this case also $\gamma=0$.

\begin{figure}[htbp]
\centering
\includegraphics[width=18cm,height=28cm]{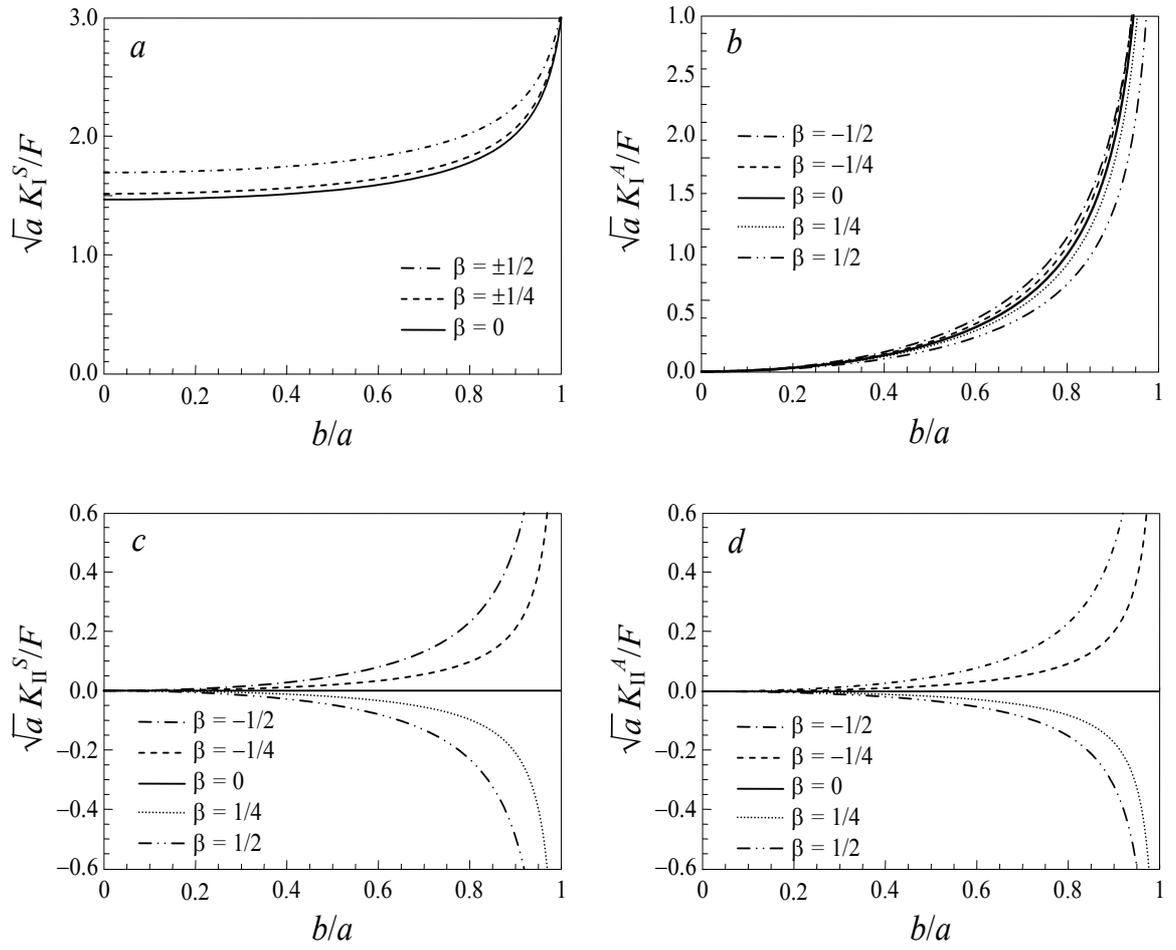}
\vspace{-14cm}
\caption[kfactor]{\footnotesize {Symmetric and anti-symmetric stress intensity factors as functions of $b/a$ computed for $\rho^{(1)}=0.74,1/\lambda^{(1)}=1, \GT^{(1)}=1/2,\rho^{(2)}=4.91,1/\lambda^{(2)}=14.3,\GT^{(2)}=2,\GF=6.4$ and different values of the Dundurs parameter $\beta$: $\beta=-1/2$,  $\beta=-1/4$, $\beta=0$, $\beta=1/4$, $\beta=1/2$.}}
\label{kfactor}
\end{figure}

 The complex stress intensity factor has been computed for $\rho^{(1)}=0.74,1/\lambda^{(1)}=1, \GT^{(1)}=1/2,\rho^{(2)}=4.91,\GF=6.4$ (here we have considered that material $(1)$ is alluminium and material $(2)$ is boron \citep{Suo2}), $\GT^{(1)}=1/2, \GT^{(2)}=2$ and five different values of the Dundurs oscillation parameter $\beta=\left\{-1/4, -1/2, 0, 1/4, 1/2\right\}$. The values have been normalized multiplying by $a^{\fr{1}{2}}F^-1$ and plotted in Fig.(\ref{kfactor}) in function of the ratio $b/a$. The symmetric stress intensity factor is reported on the left of the figure, while the skew-symmetric is on the right, the real part is reported on the top while the imaginary is on the bottom. Observing the figure, we note that both the real and the imaginary part of the skew-symmetric stress intensity factor are zero for $b/a = 0$, as expected, since for $b = 0$ the loading is symmetric. As we increase $b/a$, the skew-symmetric contribution to the loading become more relevant, and the skew symmetric stress intensity factor correspondingly increases. As we can see, in the case without oscillation, corresponding to $\beta=0$, both the imaginary parts $K^{S}_{II}$ and $K^{A}_{II}$ vanish and the stress intensity factor becomes real. Both the symmetric and the skew-symmetric stress intensity factors diverge for $b/a\rightarrow 0$, because a point force is approaching the crack tip.

\begin{figure}[htbp]
\centering
\includegraphics[width=18cm,height=28cm]{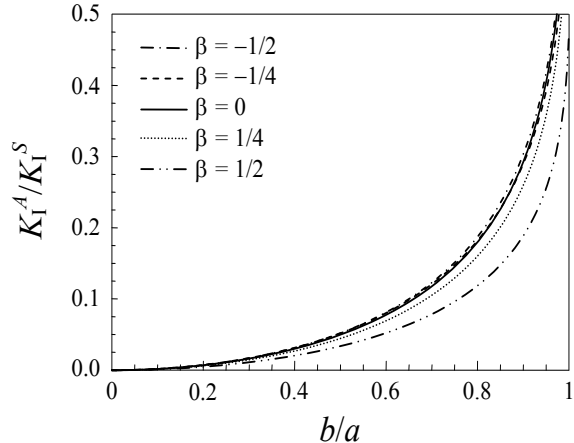}
\vspace{-20cm}
\caption[kfacratios]{\footnotesize {Ratio $K_{I}^{A}/K_{I}^{S}$ computed for $\rho^{(1)}=0.74,1/\lambda^{(1)}=1, \GT^{(1)}=1/2,\rho^{(2)}=4.91,1/\lambda^{(2)}=14.3,\GT^{(2)}=2,\GF=6.4$ and different values of the Dundurs parameter $\beta$: $\beta=-1/2$,  $\beta=-1/4$, $\beta=0$, $\beta=1/4$, $\beta=1/2$ reported in function of $b/a$.}}
\label{kfacratios}
\end{figure}

 In order to characterize the magnitude of the skew-symmetric stress intensity factor respect to the symmetric stress intensity factor, the ratio $K^{A}_{I}/K^{S}_{I}$ is 
 %reported in 
 evaluated as  a function of $b/a$ in Fig.(\ref{kfacratios}). We observe that as $b/a$ increases, $K^{A}_{I}$ may reach the $40\%$ of $K^{S}_{I}$, as a consequence we can say that the contribution of the skew-symmetric part of the loading is not negligible, and needs to be taken into account in perturbative analysis of interfacial cracks between two dissimilar anisotropic elastic materials subject to asymmetric forces systems applied on the crack faces \citep{PiccMish4,PiccMish1}. \\
 %The analytical method developed and illustrated in the article makes possible the evaluation
%of the coefficients of the higher order terms of the asymptotic expressions of the physical fields, and according to the results reported in reference \cite{PiccMish1} for the case of isotropic elastic media, also for anisotropic materials we expect that the higher order asymptotics terms are even more sensitive to the asymmetric part of the loading
\vspace{2cm}
\vspace{-4cm}
\newpage

\section{Conclusions}
The developed general approach for the derivation of symmetric and skew-symmetric weight function matrices for interfacial plane cracks between dissimilar anisotropic materials, based on Stroh formulation of displacements and stress fields, have been discussed in details and tested by means of the application to the case of a crack placed at the interface between two orthotropic materials under plane stress. The skew-symmetric weight functions obtained by means of the proposed method have been compared to those obtained for the same problem by the construction of the full-field singular solution of the elasticity problem in a half-plane \citep{PiccMish1}, the comparison between the two different solutions is reported in Appendix A. The perfect equality detected between the expressions derived by means of two distinct approaches is an important ulterior proof for the obtained results. 

Since the proposed Stroh representation is valid for stationary and steady-state elasticity problems in many anisotropic media \citep{Stroh1,Suo1,Ting1}, it can be utilized for evaluating explicit weight functions for plane interfacial cracks in several kind of materials (Stroh analysis have been proposed for example in quasi-crystals \citep{RadMar1}, piezoelectrics \citep{SuoKuo1} and poroelastics media \citep{GauKel1}. The derived weight functions can be used in many important applications: in perturbative expansions for growing cracks or wavy cracks problems \citep{PiccMish2}, and in the computation of the stress intensity factor for non-symmetric self-balanced load generated by a system of point-forces applied on the crack faces. An example of stress intensity factor evaluation for an asymmetric loading is reported in Section (\ref{example}): weight functions matrices obtained for orthotropic bi-materials have been used in the computation, and the results show that the contribution of the skew-symmetric part of the loading is not negligible and must be considered in the asymptotic expressions of the stress near the crack tip, as it has already been demonstrated for the case of isotropic media \citep{PiccMish1,PiccMish4}.
\section{Acknowledgements}
The authors are grateful to Prof. G. Mishuris and Dr. A. Piccolroaz for the fruitful discussions about weight functions theory and their valuable suggestions. L. M. and E. R. gratefully acknowledge financial support from the ''Cassa di Risparmio di Modena'' in the framework of the International Research Project $2009-2010$ "Modelling of crack propagation in complex materials".

\section*{Appendix A}
In this appendix, the Fourier transforms of the singular solution of the interfacial crack problem between two dissimilar orthotropic materials are derived by solving a boundary value problem for a semi-infinite half-plane subjected to traction boundary conditions at its boundary, following the procedure illustrated in \citet{PiccMish1}. The derived expressions for the singular displacements and for the symmetric and skew-symmetric weight functions are compared to those obtained in Section (\ref{weight_func}) by means of the direct solution of the Riemann-Hilbert problem \eq{eq5}. The perfect agreement detected between the expressions derived using two different approaches is an important test for the obtained results.

Initially, we consider the lower half-plane, denoted in the article by the superscript $^{(2)}$. Introducing the Fourier transform of the stresses respect to the variable $x_{1}$, we consider the component $\hat{\Gs}_{22}$ as the primary unknown function, so that the plane strain elasticity problem for the orthotropic material $^{(2)}$ is reduce to the following ordinary differential equation:
\beq
s_{11}^{(2)}\hat{\Gs}_{22}^{-''''}-\xi^{2}\left(s_{66}^{(2)}+2s_{12}^{(2)}\right)\hat{\Gs}_{22}^{-''}+\xi^{4}s_{22}^{(2)}\hat{\Gs}_{22}^{-}=0
\eequ{ODE}
Where a prime denotes the derivatives respect to $x_{2}$. The characteristic equation associated to \eq{ODE} is:
$$
[\Go^{(2)}]^{4}s_{11}^{(2)}-\xi^{2}\left(s_{66}^{(2)}+2s_{12}^{(2)}\right)[\Go^{(2)}]^{2}+\xi^{4}s_{22}^{(2)}=0, \quad \xi\in\textrm{R}
$$
Introducing $\nu^{(2)}=\Go^{(2)}/|\xi|$ and using the same notation of section \ref{ortho}, this characteristic equation becomes:
\beq
\Gl^{(2)}[\nu^{(2)}]^{4}-2\rho^{(2)}(\Gl^{(2)})^{\fr{1}{2}}[\nu^{(2)}]^{2}+1=0
\eequ{ODE_char}
Assuming that this equation possesses four distinct roots, ($\rho^{(2)}\neq 1$), the general solution of the plane strain elasticity problem in the lower half-plane is:
\beq
\hat{\Gs}_{22}^{-}(\xi,x_2)=A^{(2)}_{1}e^{|\xi|\nu^{(2)}_{1}x_2}+A^{(2)}_{2}e^{|\xi|\nu^{(2)}_{2}x_2},\quad\hat{\Gs}_{11}^{-}(\xi,x_2)=-\fr{1}{\xi^{2}}\hat{\Gs}_{22}^{-''},\quad\hat{\Gs}_{21}^{-}(\xi,x_2)=-\fr{i}{\xi}\hat{\Gs}^{-'}_{22},
\eequ{stresses-}
The Fourier transform on the displacements components are:
\beq
\hat{u}_{1}^{-}=\fr{i}{\xi}\left(s_{11}^{(2)}\hat{\Gs}_{11}^{-}+s_{12}^{(2)}\hat{\Gs}_{22}^{-}\right),\quad \hat{u}_{2}^{-}=\fr{1}{\xi^{2}}\left(s_{11}^{(2)}\hat{\Gs}_{11}^{-'}+s_{12}^{(2)}\hat{\Gs}_{22}^{-'}\right),
\eequ{displ-}
Where only the two eigenvalues with positive real part ($\textrm{Re}^{(2)}\nu>0$), such that the stresses vanish at the infinity ($\hat{\Gs}_{ij}^{-}\rightarrow0\quad\textrm{for}\quad x_2\rightarrow -\infty$), has been accounted. Remembering the conditions for having positive definetess of the strain energy density introduced in section \ref{ortho}, the two eigenvalues with positive real part become:
\begin{eqnarray}
\nu_{1}^{(2)} & = & \left[\lambda^{-\fr{1}{4}}(n+m)\right]^{(2)},\quad \nu_{2}^{(2)}=\left[\lambda^{-\fr{1}{4}}(n-m)\right]^{(2)},\quad\textrm{for}\quad 1 <\rho^{(2)}<+\infty,\nonumber\\
\nu_{1}^{(2)} & = & \left[\lambda^{-\fr{1}{4}}(n+im)\right]^{(2)},\quad \nu_{2}^{(2)}=\left[\lambda^{-\fr{1}{4}}(n-im)\right]^{(2)},\quad\textrm{for}\quad -1 <\rho^{(2)}<1,\nonumber
\end{eqnarray} 
From this form it is straightforward to note that these eigenvalues can be expressed in function of Stroh eigenvalues introduced in second section by means of the relation:
\beq
\nu_{1}^{(2)}=-i\mu_{1}^{(2)}, \qquad \nu_{2}^{(2)}=-i\mu_{2}^{(2)},
\eequ{etamu} 
In order to derive the weight functions, we need to evaluate explicit expressions for the singular displacements utilizing the \eq{displ-}. The boundary conditions along the boundary $x_{2}=0^{-}$ are defined as follows:      
$$
\hat{\Gs}_{22}^{-}(\xi,x_{2}=0^{-})=\hat{\GS}_{2}^{-}(\xi),\quad \hat{\Gs}_{21}^{-}(\xi,x_2=0^{-})=\hat{\GS}_{1}^{-}(\xi),\quad \xi\in\textrm{R}
$$  
where $\GS_{21}^{-},\GS_{22}^{-}$ are the components of the singular traction defined in section \ref{ortho}, (equations \eq{trac_sing1} and \eq{trac_sing2}). It follows that:
\begin{eqnarray}
\hat{\Gs}_{22}^{-}(\xi,x_{2}=0^{-}) & = & A^{(2)}_{1}+A^{(2)}_{2}=\hat{\GS}_{2}^{-}(\xi),\nonumber\\
\hat{\Gs}_{21}^{-}(\xi,x_{2}=0^{-}) & = & -i\mbox{sign}(\xi)\left(A^{(2)}_{1}\nu_{1}^{(2)}+A^{(2)}_{2}\nu_{2}^{(2)}\right)=\hat{\GS}_{1}^{-}(\xi),\nonumber
\end{eqnarray}
and thus:
\begin{eqnarray}
A^{(2)}_{1} & = & \fr{\nu_{2}^{(2)}\hat{\GS}_{2}^{-}-i\mbox{sign}(\xi)\hat{\GS}_{1}^{-}}{\nu_{2}^{(2)}-\nu_{1}^{(2)}},\\
A^{(2)}_{2} & = & \fr{i\mbox{sign}(\xi)\hat{\GS}_{1}^{-}-\nu_{2}^{(2)}\hat{\GS}_{2}^{-}}{\nu_{1}^{(2)}-\nu_{1}^{(2)}},
\end{eqnarray}
The Fourier transforms of the singular displacements fields are then:\\

$
\hat{u}_{1}^{-}(\xi,x_{2})=
$
$$
=\fr{1}{\xi(\nu_{2}^{(2)}-\nu_{1}^{(2)})}\left\{\left[\mbox{sign}(\xi)\left(s_{12}^{(2)}-s_{11}^{(2)}\left[\nu_{1}^{(2)}\right]^{2}\right)\hat{\GS}_{1}^{-}+i\nu_{2}^{(2)}\left(s_{12}^{(2)}-s_{11}^{(2)}\left[\nu_{1}^{(2)}\right]^{2}\right)\hat{\GS}_{2}^{-}\right]e^{|\xi|\nu^{(2)}_{1}x_{2}}+\right.
$$
\beq
\left.+\left[\mbox{sign}(\xi)\left(s_{11}^{(2)}\left[\nu_{2}^{(2)}\right]^{2}-s_{12}^{(2)}\right)\hat{\GS}_{1}^{-}+i\nu_{1}^{(2)}\left(s_{11}^{(2)}\left[\nu_{2}^{(2)}\right]^{2}-s_{12}^{(2)}\right)\hat{\GS}_{2}^{-}\right]e^{|\xi|\nu^{(2)}_{2}x_{2}}\right\}
\eequ{u1-sing}

$
\hat{u}_{2}^{-}(\xi,x_{2})=
$
$$
=\fr{1}{\xi(\nu_{2}^{(2)}-\nu_{1}^{(2)})}\left\{\left[i\left(s_{11}^{(2)}\left[\nu_{1}^{(2)}\right]^{2}-s_{12}^{(2)}-s_{66}^{(2)}\right)\nu_{1}^{(2)}\hat{\GS}_{1}^{-}+\right.\right.
$$
$$
\left.+\mbox{sign}(\xi)\left(s_{12}^{(2)}+s_{66}^{(2)}-s_{11}^{(2)}\left[\nu_{1}^{(2)}\right]^{2}\right)\nu_{1}^{(2)}\nu_{2}^{(2)}\hat{\GS}_{2}^{-}\right]e^{|\xi|\nu^{(2)}_{1}x_{2}}+\left[i\left(s_{12}^{(2)}+s_{66}^{(2)}-s_{11}^{(2)}\left[\nu_{2}^{(2)}\right]^{2}\right)\nu_{2}^{(2)}\hat{\GS}_{1}^{-}+\right.
$$
\beq
\left.\left.+\mbox{sign}(\xi)\left(s_{11}^{(2)}\left[\nu_{2}^{(2)}\right]^{2}-s_{12}^{(2)}-s_{66}^{(2)}\right)\nu_{1}^{(2)}\nu_{2}^{(2)}\hat{\GS}_{2}^{-}\right]e^{|\xi|\nu^{(2)}_{2}x_{2}}\right\}
\eequ{u2-sing}

For the upper half-plane, we find the same expressions, subject to replacing $|\xi|$ with $-|\xi|$ and the superscript $(2)$ with $(1)$ \cite{PiccMish1}. From equations \eq{u1-sing} and \eq{u2-sing} and their corresponding expressions on the upper half-plane, we can derive the traces of the singular displacements transforms 
on the plane containing the crack:
\begin{eqnarray}
\hat{U}^{+}_{1}(\xi) & = & \hat{u}^{+}_{1}(\xi,x_{2}=0^{+})=\left(-\fr{[(\nu_{2}+\nu_{1})s_{11}]^{(1)}}{|\xi|},i\fr{[s_{12}+s_{11}\nu_{1}\nu_{2}]^{(1)}}{\xi}\right)\hat{\BGS}^{-}\nonumber\\
\hat{U}^{+}_{2}(\xi) & = & \hat{u}^{+}_{2}(\xi,x_{2}=0^{+})=\left(i\fr{[(s_{11}+s_{66})-(\nu_{2}+\nu_{1}+\nu_{1}\nu_{2})s_{11}]^{(1)}}{\xi},-\fr{[s_{11}\nu_{1}\nu_{2}(\nu_{2}+\nu_{1})]^{(1)}}{|\xi|}\right)\hat{\BGS}^{-}\nonumber\\
\hat{U}^{-}_{1}(\xi) & = & \hat{u}^{-}_{1}(\xi,x_{2}=0^{-})=\left(\fr{[(\nu_{2}+\nu_{1})s_{11}]^{(2)}}{|\xi|},i\fr{[s_{12}+s_{11}\nu_{1}\nu_{2}]^{(2)}}{\xi}\right)\hat{\BGS}^{-}\nonumber\\
\hat{U}^{-}_{2}(\xi) & = & \hat{u}^{-}_{2}(\xi,x_{2}=0^{-})=\left(i\fr{[(s_{11}+s_{66})-(\nu_{2}+\nu_{1}+\nu_{1}\nu_{2})s_{11}]^{(2)}}{\xi},\fr{[s_{11}\nu_{1}\nu_{2}(\nu_{2}+\nu_{1})]^{(2)}}{|\xi|}\right)\hat{\BGS}^{-}\nonumber
\end{eqnarray}
Using the relation \eq{etamu} between $\eta_{1,2}$ and the Stroh eigenvalues:
$$
\nu_{1}=-i\mu_{1}, \qquad \nu_{2}=-i\mu_{2},
$$
and considering the following relations
$$
\nu_{1}^{2}+\nu_{2}^{2}=-(\mu_{1}^{2}+\mu_{2}^{2})=\fr{2s_{12}+s_{66}}{s_{11}},\quad \nu_{1}^{2}\nu_{2}^{2}=\mu_{1}^{2}\mu_{2}^{2}=\fr{s_{22}}{s_{11}};
$$
we deduce the expressions for singular displacements along the axes of propagation of the crack ($x_2=0$),  
%can be rewritten 
as follows:
\begin{eqnarray}
\hat{U}^{+}_{1}(\xi) & = & \hat{u}^{+}_{1}(\xi,x_{2}=0^{+})=\left(-\fr{[\mbox{Im}(\mu_{1}+\mu_{2})s_{11}]^{(1)}}{|\xi|},-i\fr{[s_{11}\mu_{1}\mu_{2}-s_{12}]^{(1)}}{\xi}\right)\hat{\BGS}^{-}\nonumber\\
\hat{U}^{+}_{2}(\xi) & = & \hat{u}^{+}_{2}(\xi,x_{2}=0^{+})=\left(i\fr{[s_{11}\overline{\mu_{1}} \overline{\mu_{2}}-s_{12}]^{(1)}}{\xi},\fr{1}{|\xi|}\left[\mbox{Im}\left(\fr{1}{\mu_{1}}+\fr{1}{\mu_{2}}\right)\right]^{(1)}\right)\hat{\BGS}^{-}\nonumber\\
\hat{U}^{-}_{1}(\xi) & = & \hat{u}^{-}_{1}(\xi,x_{2}=0^{-})=\left(\fr{[\mbox{Im}(\mu_{1}+\mu_{2})s_{11}]^{(2)}}{|\xi|},-i\fr{[s_{11}\mu_{1}\mu_{2}-s_{12}]^{(2)}}{\xi}\right)\hat{\BGS}^{-}\nonumber\\
\hat{U}^{-}_{2}(\xi) & = & \hat{u}^{-}_{2}(\xi,x_{2}=0^{-})=\left(i\fr{[s_{11}\overline{\mu_{1}} \overline{\mu_{2}}-s_{12}]^{(2)}}{\xi},-\fr{1}{|\xi|}\left[\mbox{Im}\left(\fr{1}{\mu_{1}}+\fr{1}{\mu_{2}}\right)\right]^{(2)}\right)\hat{\BGS}^{-}\nonumber\\
\end{eqnarray}
These expressions can be written in the same form of the physical displacements \eq{u+} and \eq{u-}:
\begin{eqnarray}
\hat{\BU}^+ (\xi) & = & \Big\{  \fr{1}{2 \xi} (\BY^{(1)} - \ov{\BY}^{(1)}) - \fr{1}{2 |\xi |}     (  \BY^{(1)} + \ov{\BY}^{(1)}  )   \Big\} \hat{\BGS}\\
\hat{\BU}^- (\xi) & = & \Big\{  \fr{1}{2 \xi} (\BY^{(2)} - \ov{\BY}^{(2)}) + \fr{1}{2 |\xi |}     (  \BY^{(2)} + \ov{\BY}^{(2)}  )   \Big\} \hat{\BGS} 
\end{eqnarray}
where the hermitian matrices $\BY^{(1)}$ and $\BY^{(2)}$ possess exactly the same form  \eq{Ymatrix}, evaluated by specializing the Stroh formalism to the case of a two-dimensional orthotropic material \citep{Suo1}:
\beq
\BY^{(1),(2)} = \pmatrix{\left[s_{11}\textrm{Im}(\mu_{1}+\mu_{2})\right]^{(1),(2)} & -i\left[s_{11}\mu_{1}\mu_{2}-s_{12}\right]^{(1),(2)} \cr i\left[s_{11}\overline{\mu}_{1}\overline{\mu}_{2}-s_{12}\right]^{(1),(2)}  & -\left[s_{22}\textrm{Im}\left(\fr{1}{\mu_{1}}+\fr{1}{\mu_{2}}\right)\right]^{(1),(2)}}
\eequ{Ymatrix2}
%As in the second section, the 
The Fourier transforms of the symmetric and skew-symmetric weight functions are defined respectively as the jump and the average of the singular displacements across the plane containing the crack:
\begin{eqnarray}
[\hat{\BU}](\xi)                & = & \hat{\BU}^+ (\xi)-\hat{\BU}^- (\xi) 
=\fr{1}{|\xi|} \Big\{ i ~ \mbox{sign} (\xi) ~ \mbox{Im} (\BY^{(1)}- \BY^{(2)})  - \mbox{Re} (\BY^{(1)} + \BY^{(2)})   \Big\} \hat{\BGS} \nonumber\\
\langle \hat{\BU} \rangle (\xi) & = & \fr{1}{2}\left(\hat{\BU}^+ (\xi)+\hat{\BU}^- (\xi)\right) = \fr{1}{2 |  \xi |} \Big\{ i ~ \mbox{sign} (\xi) ~ \mbox{Im} (\BY^{(1)}+ \BY^{(2)})  - \mbox{Re} (\BY^{(1)} - \BY^{(2)})   \Big\} \hat{\BGS} \nonumber\\
\end{eqnarray}
We have finally recovered the expressions \eq{eq6} and \eq{eq7}, previously derived from the direct solution of the Riemann-Hilbert problem for the interfacial crack by means of the Stroh formalism, consequently, we can say that the two alternative formulations are perfectly equivalent, and that our result is proved by this further test. 
\include{bibliografia} % the bibliography (of the thesis)
\bibliography{WF_bib} % bibliografia con bibtex

\begin{thebibliography}{27}
\expandafter\ifx\csname natexlab\endcsname\relax\def\natexlab#1{#1}\fi
\expandafter\ifx\csname url\endcsname\relax
  \def\url#1{{\tt #1}}\fi
\expandafter\ifx\csname urlprefix\endcsname\relax\def\urlprefix{URL }\fi

\bibitem[{Antipov(1999)}]{Ant1}
Antipov, Y.~A. (1999).
\newblock An exact solution of the 3-{D} problem of an interface semi-infinite
  plane crack.
\newblock {\em J. Mech. Phys. Solids\/}, {\em 47\/}, 1051--1093.

\bibitem[{Bercial-Velez et~al.(2005)Bercial-Velez, Antipov, \&
  Movchan}]{BercVel1}
Bercial-Velez, J.~P., Antipov, Y.~A., \& Movchan, A.~B. (2005).
\newblock High-order asymptotics and perturbation problems for 3{D} interfacial
  cracks.
\newblock {\em J. Mech. Phys. Solids\/}, {\em 53\/}, 1128--1162.

\bibitem[{Bueckner(1985)}]{Bueck1}
Bueckner, H.~F. (1985).
\newblock Weight functions and fundamental fields for the penny-shaped and the
  half plane crack in three-space.
\newblock {\em Int. J. Solids Struct.\/}, {\em 23\/}, 57--93.

\bibitem[{Bueckner(1989)}]{Bueck2}
Bueckner, H.~F. (1989).
\newblock Observations on weight functions.
\newblock {\em Eng. Anal. Bound. Elem.\/}, {\em 6\/}, 3--18.

\bibitem[{Dundurs(1969)}]{Dund1}
Dundurs, J. (1969).
\newblock Discussion of a paper by {D}. {B}. {B}ogy.
\newblock {\em J. Appl. Mech.\/}, {\em 36\/}, 650--652.

\bibitem[{Gao(1991)}]{Gao2}
Gao, H. (1991).
\newblock Weight function analysis for interface cracks: mis-mach versus
  oscillation.
\newblock {\em J. Appl. Mech.\/}, {\em 58\/}, 931--938.

\bibitem[{Gao(1992)}]{Gao1}
Gao, H. (1992).
\newblock Weight function method for interfacial cracks in anisotropic
  bimaterials.
\newblock {\em Int. J. Fract.\/}, {\em 56\/}, 139--158.

\bibitem[{Gao et~al.(1992)Gao, Abbudi, \& Barnett}]{GaoAbbu1}
Gao, H., Abbudi, M., \& Barnett, D.~M. (1992).
\newblock Interfacial crack-tip field in anisotropic elastic solids.
\newblock {\em J. Mech. Phys. Solids\/}, {\em 40\/}, 393--416.

\bibitem[{Gautier et~al.(2011)Gautier, Kelders, Groby, Dazel, Ryck, \&
  Leclaire}]{GauKel1}
Gautier, G., Kelders, L., Groby, J.~P., Dazel, O., Ryck, L.~D., \& Leclaire, P.
  (2011).
\newblock Propagation of acoustic waves in a one-dimensional macroscopically
  inhomogeneous poroelastic material.
\newblock {\em J. Acoust. Soc. Am.\/}, {\em 130\/}, 1390--1398.

\bibitem[{Gupta et~al.(1992)Gupta, Argon, \& Suo}]{GupArg1}
Gupta, V., Argon, A.~S., \& Suo, Z. (1992).
\newblock Crack deflection at an interface between two orthotropic media.
\newblock {\em J. Appl. Mech.\/}, {\em 59\/}, S79--S87.

\bibitem[{Hwu(1993)}]{Hwu1}
Hwu, C. (1993).
\newblock Relations among {S}troh, {L}ekhnitskii and {M}uskhelishvili
  formulations.
\newblock In {\em Proc. of the 10th National Conference on Mechanical
  Engineering\/}, (pp. 517--525). Taipei, Taiwan, R.O.C:.

\bibitem[{Lazarus \& Leblond(1998)}]{LazLebl1}
Lazarus, V., \& Leblond, J.~B. (1998).
\newblock Three-dimensional crack-face weight functions for the semi-infinite
  interface crack-i: variation of the stress intensity factors due to some
  small perturbation of the crack front.
\newblock {\em J. Mech. Phys. Solids\/}, {\em 46\/}, 489--511.

\bibitem[{Lekhnitskii(1963)}]{Lekh1}
Lekhnitskii, S.~G. (1963).
\newblock {\em Theory of elasticity of an anisotropic mody\/}.
\newblock Holden-{D}ay, San Francisco.

\bibitem[{Ma \& Chen(2004)}]{MaChen1}
Ma, L., \& Chen, Y. (2004).
\newblock Weight functions for interface cracks in dissimilar anisotropic
  materials.
\newblock {\em Acta Mech. Sin.\/}, {\em 20\/}, 82--88.

\bibitem[{Meade \& Keer(1984)}]{MeadKeer1}
Meade, K.~P., \& Keer, L.~M. (1984).
\newblock On the problem of a pair of point forces applied to the faces of a
  semi-infinite plane crack.
\newblock {\em J. Elasticity\/}, {\em 14\/}, 3--14.

\bibitem[{Mishuris \& Kuhn(2001)}]{MishKuhn1}
Mishuris, G.~S., \& Kuhn, G. (2001).
\newblock Asymptotic behaviour of the elastic solution near the tip of a crack
  situated at a nonideal interface.
\newblock {\em Z. Angew. Math. Mech.\/}, {\em 81\/}, 811--826.

\bibitem[{Piccolroaz \& Mishuris(2011)}]{PiccMish3}
Piccolroaz, A., \& Mishuris, G. (2011).
\newblock Integral identities for a semi-infinite interfacial crack in 2{D} and
  3{D} elasticity.
\newblock {\em Ar{X}iv:1110.3612v1\/}.

\bibitem[{Piccolroaz et~al.(2007)Piccolroaz, Mishuris, \& Movchan}]{PiccMish2}
Piccolroaz, A., Mishuris, G., \& Movchan, A.~B. (2007).
\newblock Evaluation of the {L}azarus-{L}eblond constants in the asymptotic
  model for the interfacial wavy crack.
\newblock {\em J. Mech. Phys. Solids\/}, {\em 55\/}, 1575--1600.

\bibitem[{Piccolroaz et~al.(2009)Piccolroaz, Mishuris, \& Movchan}]{PiccMish1}
Piccolroaz, A., Mishuris, G., \& Movchan, A.~B. (2009).
\newblock Symmetric and skew symmetric weight functions in 2{D} perturbation
  models for semi-infinite interfacial cracks.
\newblock {\em J. Mech. Phys. Solids\/}, {\em 57\/}, 1657--1682.

\bibitem[{Piccolroaz et~al.(2010)Piccolroaz, Mishuris, \& Movchan}]{PiccMish4}
Piccolroaz, A., Mishuris, G., \& Movchan, A.~B. (2010).
\newblock Perturbation of mode {III} interfacial cracks.
\newblock {\em Int. J. Fract.\/}, {\em 166\/}, 41--51.

\bibitem[{Radi \& Mariano(2010)}]{RadMar1}
Radi, E., \& Mariano, P.~M. (2010).
\newblock Stationary straight cracks in quasicrystals.
\newblock {\em Int. J. Fract.\/}, {\em 166\/}, 105--120.

\bibitem[{Stroh(1962)}]{Stroh1}
Stroh, A.~N. (1962).
\newblock Steady state problems in anisotropic elasticity.
\newblock {\em J. Math. Phys.\/}, {\em 41\/}, 77--103.

\bibitem[{Suo(1990{\natexlab{a}})}]{Suo2}
Suo, Z. (1990{\natexlab{a}}).
\newblock Delamination specimens for orthotropic materials.
\newblock {\em J. Appl. Mech.\/}, {\em 57\/}, 627--634.

\bibitem[{Suo(1990{\natexlab{b}})}]{Suo1}
Suo, Z. (1990{\natexlab{b}}).
\newblock Singularities, interfaces and cracks in dissimilar anisotropic media.
\newblock {\em Proc. R. Soc. Lond. A\/}, {\em 427\/}, 331--358.

\bibitem[{Suo et~al.(1992)Suo, Kuo, Barnett, \& Willis}]{SuoKuo1}
Suo, Z., Kuo, C.-M., Barnett, D.~M., \& Willis, J.~R. (1992).
\newblock Fracture mechanics for piezoelectric ceramics.
\newblock {\em J. Mech. Phys. Solids\/}, {\em 40\/}, 739--765.

\bibitem[{Ting(1996)}]{Ting1}
Ting, T. C.~T. (1996).
\newblock {\em Anisotropic elasticity: theory and applications\/}.
\newblock Oxford {U}niversity {P}ress.

\bibitem[{Willis \& Movchan(1995)}]{Wilmov1}
Willis, J.~R., \& Movchan, A.~B. (1995).
\newblock Dynamic weight function for a moving crack. {I}. mode {I} loading.
\newblock {\em J. Mech. Phys. Solids\/}, {\em 43\/}, 319--341.

\end{thebibliography}
%\addcontentsline{toc}{chapter}{bibliografia}
\bibliographystyle{apa-good}
% Bibliographic references with the natbib package:
% Parenthetical: \citep{Bai92} produces (Bailyn 1992).
% Textual: \citet{Bai95} produces Bailyn et al. (1995).
% An affix and part of a reference:
%   \citep[e.g.][Ch. 2]{Bar76}
%   produces (e.g. Barnes et al. 1976, Ch. 2).
% \bibitem[Names(Year)]{label} or \bibitem[Names(Year)Long names]{label}.
% (\harvarditem{Name}{Year}{label} is also supported.)
% Text of bibliographic item

%\bibitem[]{
\end{document}